\documentclass[aps,twocolumn,floats,prd,nofootinbib,,superscriptaddress,letterpaper]{revtex4} 
\usepackage[dvips]{graphicx} %
\usepackage{epsfig,amsmath}
\usepackage{amssymb}
\usepackage{rotate}
\usepackage{color,comment}
\usepackage{bm}

\DeclareFontFamily{OT1}{pzc}{}
\DeclareFontShape{OT1}{pzc}{m}{it}%
            {<-> s * [1.10] pzcmi7t}{}
\DeclareMathAlphabet{\mathscr}{OT1}{pzc}%
                                {m}{it}

\newcommand{\be}{\begin{eqnarray}}
\newcommand{\ee}{\end{eqnarray}}
\newcommand{\bea}{\begin{eqnarray}}
\newcommand{\eea}{\end{eqnarray}}
\def\ba#1\ea{\begin{align}#1\end{align}}

\newcommand{\refeq}[1]{Eq.~(\ref{eq:#1})}          
\newcommand{\refeqs}[2]{Eqs.~(\ref{eq:#1})--(\ref{eq:#2})}          
          
\newcommand{\reffig}[1]{Fig.~\ref{fig:#1}}

\newcommand{\vs}{\nonumber\\}       
\newcommand{\refsec}[1]{Sec.~\ref{sec:#1}}          
\newcommand{\refapp}[1]{App.~\ref{app:#1}}          
 
\newcommand{\chg}[1]{#1}

\renewcommand{\v}[1]{\mathbf{#1}}

\newcommand{\vx}{\v{x}}

\newcommand{\vk}{\v{k}}

\renewcommand{\vr}{\v{r}}

\newcommand{\ex}[1]{\langle #1 \rangle}

\newcommand{\<}{\langle}
\renewcommand{\>}{\rangle}

\renewcommand{\k}{\kappa}

\newcommand{\nhat}{\hat{n}}
\newcommand{\vnhat}{\v{\hat{n}}}

\renewcommand{\l}{\ell}

\renewcommand{\d}{\delta}
\renewcommand{\t}{\vartheta}

\newcommand{\g}{\gamma}
\newcommand{\D}{\Delta}

\newcommand{\rhob}{\overline{\rho}}

\newcommand{\vl}{\bm{\l}}

\newcommand{\iMpch}{\,h~{\rm Mpc}^{-1}}

\newcommand{\s}{\sigma}

\newcommand{\M}{\mathcal{M}}
\newcommand{\T}{\mathcal{T}}
\newcommand{\C}{\mathcal{C}}

\newcommand{\A}{\mathcal{A}}
\newcommand{\B}{\mathcal{B}}
\renewcommand{\O}{\mathcal{O}}
\renewcommand{\o}{\rho}

\renewcommand{\P}{\mathcal{P}}

\newcommand{\fNL}{f_{\rm NL}}

\def\fnc{$\mathrm{FNC}$}
\def\fncb{$\overline{\mathrm{FNC}}$}
\def\Pb{\bar P}

\def\chis{\chi_*}

\def\arrowlength{1cm}

\begin{document}

\title{The Observed Squeezed Limit of Cosmological Three-Point Functions}

\author{Enrico Pajer}
\affiliation{Department of Physics, Princeton University, Princeton, NJ 08544, USA}

\author{Fabian Schmidt}
\affiliation{Department of Astrophysical Sciences, Princeton University, Princeton, NJ 08544, USA}
\affiliation{Einstein fellow}

\author{Matias Zaldarriaga}
\affiliation{Institute for Advanced Study, Princeton, NJ 08544, USA}

\begin{abstract}

The squeezed limit of the three-point function of cosmological perturbations is a powerful discriminant of different models of the early Universe. We present a conceptually simple and complete framework to relate any primordial bispectrum in this limit to late time observables, such as the CMB temperature bispectrum and the scale-dependent halo bias.  We employ a series of convenient coordinate transformations to capture the leading non-linear effects of cosmological perturbation theory on these observables. This makes crucial use of Fermi Normal Coordinates and their conformal generalization, which we introduce here and discuss in detail. As an example, we apply our formalism to standard slow-roll single-field inflation.  We show explicitly that Maldacena's results for the squeezed limits of the scalar bispectrum [proportional to $(n_{s}-1)$ in comoving gauge] and the tensor-scalar-scalar bispectrum lead to no deviations from a Gaussian universe, except for projection effects. In particular, the primordial contributions to the squeezed CMB bispectrum and scale dependent halo bias vanish, and there are no \chg{primordial} ``fossil'' correlations between long-wavelength tensor perturbations and small-scale perturbations.  The contributions to observed correlations are then only due to projection effects such as gravitational lensing and redshift perturbations.
\end{abstract}
\date{\today}

\maketitle


\section{Introduction}
\label{sec:intro}

\begin{figure*}
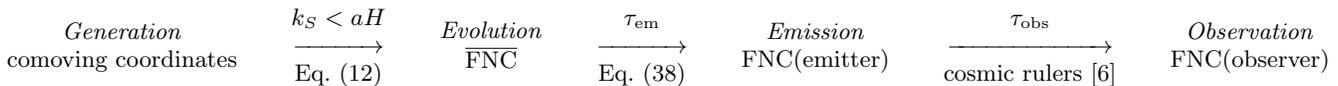

\begin{tabular}{ccccccccc}
 \begin{tabular}{c} \emph{Generation}\\ comoving coordinates\end{tabular} \quad & \quad  \begin{tabular}{c} $k_{S}<aH$\\ $\xrightarrow{\hspace*{\arrowlength}}$ \\ \refeq{FNCbtrans} \end{tabular} \quad & \quad \begin{tabular}{c} \emph{Evolution}\\ \fncb\end{tabular} \quad&\quad  \begin{tabular}{c} $\tau_{\rm em}$ \\ $\xrightarrow{\hspace*{\arrowlength}}$ \\ \refeq{fncbtofnc} \end{tabular} \quad& \quad   \begin{tabular}{c} \emph{Emission}\\ \fnc (emitter) \end{tabular} \quad&\quad \begin{tabular}{c} $\tau_{\rm obs}$ \\ $\xrightarrow{\hspace*{2cm}}$ \\ cosmic rulers~\cite{stdruler} \end{tabular} \quad&\quad \begin{tabular}{c}  \emph{Observation} \\ \fnc (observer)\end{tabular} 
\end{tabular}
\caption{Sequence of coordinates employed in the computation of observables predicted by a primordial bispectrum in the squeezed limit. The arrows represent the change from one set of coordinates to the next, indicating when this transformation is most conveniently performed and with a reference to the relevant equations. $\tau_{\rm em}$ and $\tau_{\rm obs}$ denote the conformal times at which the observed photons were emitted and observed, respectively. \label{fig:overview}}
\end{figure*}

In this paper we present a simple and complete framework to derive the late-time physical observables produced by the squeezed limit of primordial three-point functions, in particular those produced during inflation.  
These observables include the squeezed limit of the CMB bispectrum and the scale-dependent bias for large-scale structure tracers which crucially depends on the bispectrum in the squeezed limit \cite{dalal/etal:2008,verde/matarrese:2008,schmidt/kamionkowski:2010}. The main obstacle one encounters is the fact that, in order to calculate the bispectrum consistently, one has to work to second order in perturbation theory throughout. We show how this obstacle is overcome when one is interested in the squeezed limit. We take advantage of the fact that different sets of coordinates are convenient in different cosmological epochs and there is a careful choice that makes the computation easy and transparent.  In the remainder of the introduction, we present a convenient way to choose coordinates by discussing the different steps that are summarized in \reffig{overview}.

A first step consists in computing the primordial correlators generated during the phase of \textit{primordial inflation}. For this purpose, coordinates are used that are comoving with respect to some unperturbed universe (see e.g.~\cite{maldacena:2003}).  A second step consists in describing the local \textit{late-time physical processes}, such as for example the decoupling of photons during recombination or the formation of a dark matter halo, that lead to the generation of some light signal that eventually will reach the Earth. Since these processes take place in regions that are smaller in size than the curvature of the background metric (horizon scale $H^{-1}$), they are best described in \textit{Fermi Normal Coordinates} (FNC \cite{ManasseMisner}):  this is the unique frame (up to three Euler angles) constructed along a timelike geodesic passing through a point $P$ in which the metric is Minkowski with corrections going as the spatial distance $r_F$ from the central geodesic squared.  In an unperturbed FRW Universe, the corrections thus scale as $(H r_F)^2$, while in the presence of a metric perturbation $h_{\mu\nu}$ the corrections involve second derivatives of $h_{\mu\nu}$ (in particular, a constant or pure gradient metric perturbation is removed on all scales).  The last step consists in relating the signals from the point of emission, for example the last scattering surface, or the position of a halo, to the \textit{observables measured on Earth}, where scientists most naturally use FNC centered on the Earth's world line.

In general, calculating the evolution of the bispectrum of perturbations in any coordinate frame including modes outside the horizon requires a full second-order relativistic calculation in cosmological perturbation theory, including gravity, baryon physics, radiation transfer, etc.  This is in general a very complicated problem.  However, a drastic simplification takes place when we limit ourselves to a particular configuration of the bispectrum in which two of the momenta, say $k_{1}$ and $k_{2}$ are much larger than the third, $k_{3}$. This is often referred to as the \textit{squeezed limit} with $k_{3}\equiv k_L$ being the \textit{long mode} and $k_{1}\sim k_{2} \sim k_{S}$ being the \textit{short modes}. 

Then, as described above, we can make use of the fact that in a $k_L/k_{S}\ll 1$ expansion, the leading and next-to-leading order gravitational effects of the long-wavelength perturbation $k_L$ on the short modes can be removed by transforming to the local FNC frame, as was emphasized in \cite{BaldaufEtal,conformal,Senatore:2012ya,Senatore:2012wy}. The first non-trivial effect comes from the second spatial and time derivatives of the perturbation, which are suppressed by $\mathcal{O}(k_L^{2}/k_{S}^{2})$ (see \refsec{fnc} for a more precise statement). We will ignore terms of this order in this paper, as they correspond to actual physical interactions, which will need to be treated in detail.  However, up to this order, in the local FNC frame we can simply use linear theory. The non-Gaussian correlation will then be captured by averaging over an ensemble of FNC, each constructed to remove a different long mode.

In going from comoving coordinates to FNC, one subtlety arises from the fact that FNC are affected not only by perturbations but also by the expansion of the Universe. Because of this, no matter when the change of coordinates is performed, when using FNC we cannot avoid a period when the evolution at second order in perturbations becomes relevant.  To see this, consider an expanding universe with a single long-wavelength perturbation, in addition to short-wavelength modes as in \reffig{sketch}.  The metric in standard FNC takes a simple flat space form only on scales smaller than the Hubble scale $H^{-1}$ or the physical wavelength of the long mode $a/k_L$, whichever is smaller. Therefore for the FNC metric to be valid in a region larger than the short modes, which is what we want to describe, we need $k_{S}\gg aH$ ($k_{S}\gg k_L$ being always true by assumption). But as soon as the short modes re-enter the horizon, $k_{S}\gtrsim aH$, they start evolving, and acquire a non-trivial transfer function. In order to see that the evolution needs to be followed at non-linear order, consider our result for a general squeezed bispectrum \refeq{Bsqfinald2}.  This expression contains terms involving the derivatives with respect to scale and time of the short mode power spectrum, which in general become non-trivial when $k_{S}\gtrsim aH$ and the short modes start evolving with time.  It is clear that these terms cannot be accounted for by just multiplying each of the perturbations with the respective transfer function as \textit{linear} perturbation theory would suggest.
 
  A nice way out of this complication is to use \textit{conformal Fermi Normal Coordinates} (\fncb), in which the metric is locally in the FLRW form, i.e.~a homogeneous isotropic expanding Universe, rather than in the Minkowski form.  These coordinates are valid up to a scale $k_L^{-1}$, where $k_L$ denotes the wavenumber of the long-wavelength perturbation considered.  At late times, the small-wavelength modes are well inside the horizon and have in general been subject to non-linear evolution. Then, transforming from the \fncb~frame to the usual FNC frame at a given spacetime point simply corresponds to a rescaling of the spatial coordinates. 
  
   The final step is to relate quantities defined in the local FNC frame (e.g.~temperature of the gas, or halo mass) to the photons actually measured on Earth.  This involves the photon propagation (``projection'') effects, such as lensing and redshift-space distortions \cite{Creminelli/Zaldarriaga:2004,BoubekeurEtal,CreminelliPitrouVernizzi,Bartolo:2011wb,stdruler,Lewis:2012}.  Here, we will adopt the ``standard ruler'' approach \cite{stdruler,T} since it conveniently encompasses all projection effects for various observables.  Specifically, the derivation in \citet{stdruler} assumes that a ruler, such as a $\xi(r) = $~const contour, corresponds to a fixed physical spatial scale on a constant-proper-time hypersurface for a comoving observer, equivalent to a fixed spatial scale in FNC at emission.   

Finally, at leading order, any \emph{physical} (and necessarily non-gravitational) correlations between long-wavelength modes and small scale modes imprinted at early times and present in \fncb~then simply add to these projection effects.  The framework discussed above, which we summarize in \reffig{overview}, thus connects the bispectrum calculated in any convenient gauge during inflation with late-time observables such as the CMB bispectrum or the large-scale clustering of tracers.  It can be applied to any model, and we will give some examples for single clock inflation and the ``fossil'' scenarios studied in \cite{MasuiPen,GiddingsSloth,GiddingsSloth2,JeongKamionkowski}.  We should mention that the importance of \fnc~in applications to the calculation of inflationary perturbations has been recognized and crucially used previously in the literature, e.g.~in deriving various consistency relations \cite{Senatore:2012wy}, in proving that $\zeta$ is constant at large distances at all loop orders \cite{Senatore:2012ya} and in deriving the subleading corrections in $k_{l}/k_{s}$ to Maldacena's consistency condition \cite{conformal}. The issue of gauge artifacts in the squeezed limit of the bispectrum was also discussed in \cite{Tanaka:2011aj}.

The outline of the paper is as follows: in \refsec{sqlimit} we derive some useful formulae to transform correlators from one set of coordinates to another. In \refsec{fnc} we introduce \fncb, and show how the squeezed-limit three-point correlations calculated in the standard way transform to this frame.  In \refsec{latetime} we derive how correlations in \fncb~translate to observables, illustrating that, apart from projection effects such as lensing, the squeezed-limit three-point function in conformal Fermi coordinates is in fact the observed squeezed limit.  In other words, if squeezed-limit
correlations vanish in \fncb, then the observed squeezed limit is only due to projection effects. We conclude in \refsec{concl} and leave some technical details to the appendices. \refapp{corrgen} and \refapp{sqlimit} derive the transformation of the two and three point function in the squeezed limit, respectively, as one moves from one set of coordinates to another. In \refapp{FNC} we review the derivation of FNC and discuss their uniqueness in \refapp{FNCU}.

\section{Transformation of small-scale correlations and the squeezed bispectrum}
\label{sec:sqlimit}

In this section we present some useful formulae to transform the two-point correlation function, \refeq{xitranslin}, and three-point function and bispectrum in the squeezed limit, \refeq{Bsqfinald2}, from any one set of coordinates to another. The details of the derivation are left to appendix \refapp{corrgen} and \refapp{sqlimit}.

Consider a scalar field $\rho(x)$, where $x=(\tau,\vx)$ denotes the spacetime
position.  We work in comoving coordinates throughout, so that $x^0 = \tau$ is the conformal time.  Under a general coordinate transformation $x \to x'(x)$, the
field transforms as
\ba
\rho'(x') = \rho(x(x'))\,.
\label{eq:dtransf}
\ea
Consider a patch on a $\tau'=$~const surface centered around the position
$\vx'_0$.  We would like to derive the correlation $\xi'(\vr',\tau')$ within
that patch (defined with respect to the mean density $\rhob'(\tau')$ over the
patch) in terms of the two-point correlation function $\xi(\vr,\tau)$ of 
$\rho$ in the unprimed coordinate system.  Throughout the paper, we are
considering the case where the coordinate transformation varies slowly
over the spatial patch, that is where we can expand $x$ in terms of $x'$ as
\ba
x^{i}(x') =\:& x_0^i + A^i_{\  j} \left(x'^j - x_0'^j \right) \vs
\tau(x') =\:& \tau_0 + T_j \left(x'^j - x_0'^j \right)\,,
\label{eq:tautransf}
\ea
where $x' = (\tau', \vx')$, and
\ba
\vx_0 =\:& \vx(\vx'_0, \tau') \vs
\tau_0 =\:& \tau(\vx'_0,\tau') = \tau' + \D\tau
\vs
A^i_{\  j} =\:& \frac{\partial x^i(\vx'_0,\tau')}{\partial x'^j} 
= \d^i_{\  j} - a^i_{\  j} \,.
\ea
Note that $a^i_{\  j}$ and $\D\tau$ are functions of $\vx'_0$ and $\tau'$.  
Here we have separated $A^i_{\  j}$ into a zeroth and first order piece (the sign is chosen so that $a^i_{\  j}$ maps $x^j$ to $x'^i$ at linear order).  The
expansion to linear order in $\vx'-\vx_0'$ is sufficient, since we will only
be interested in Fourier space contributions to the coordinate transformations
up to order $k_L^2$ (see \refapp{corrgen}).  As shown in \refapp{corrgen},
the correlation function $\xi'(\vr',\tau')$ of $\rho'$ then becomes 
\ba
&\xi'(\vr',\tau') = 
\left[1 - a_{ij}\, r'^i \partial_{r'}^j + \D\tau\,\partial_\tau \right] \xi(\vr',\tau')\,.
\label{eq:xitranslin}
\ea
The dependence on $x'^j$ of the time shift \refeq{tautransf} does not contribute at leading order.

We now turn to the squeezed limit of the three-point function $\< X'(\vx'_3) \d'(\vx'_1) \d'(\vx'_2)\>$, where $\d = \d\rho/\bar\rho$ denotes the fractional perturbation to $\rho$ (the results are identical when considering $\d\rho$ instead of $\d$).  This limit corresponds to the case where $|\vx_3-\vx_1| \gg |\vx_1-\vx_2|$.  Here, $X$ stands for any other field ($X$ is not necessarily a scalar, but we suppress all tensor indices), such as for example density perturbations or tensor modes. Strictly speaking, $X$ is to be understood as coarse-grained on a scale $R > |\vx_2-\vx_1|$.  In the following, we will drop the primes on coordinates for clarity, since we will not employ the unprimed coordinates anymore.  Further, we adopt the notation $X'(\vx) = X(\vx)$, that is, the long-wavelength (coarse-grained) field $X$ is not modified under the small-scale coordinate transformation within the patch.  
In the squeezed limit, the three-point function describes the modulation of the local two-point function $\xi'(|\vx_1-\vx_2|;
\vx_0)$ at the location $\vx_0 = (\vx_1+\vx_2)/2$ by the long-wavelength field $X$ evaluated at the distant point $\vx_3$:
\ba
& \<X'(\vx_3) \d'(\vx_1) \d'(\vx_2) \> \stackrel{\rm squeezed}{=} \vs
&\hspace*{0.5cm} \< X'(\vx_3)\, \xi'(\vx_1-\vx_2;\tau)|_{X'(\vx_0)}\,\>\,.
\ea
The precise location of the point $\vx_0$ in fact does not matter in the squeezed limit (as we prove in \refapp{sqlimit}), but we have chosen the midpoint as the most natural choice.  We can now express $\<X'\d'\d'\>$ in terms of the three-point function in unprimed coordinates $\<X \d\d\>$ and contributions from the transformation of small-scale correlations [\refeq{xitranslin}].  This yields
\ba
& \<X'(\vx_3) \d'(\vx_1) \d'(\vx_2) \> \stackrel{\rm squeezed}{=} \bigg[1 - \xi_{X a_{ij}}(|\vx_3-\vx_0|) r^i \partial_{r}^j \vs
& + \xi_{X \D\tau}(|\vx_3-\vx_0|)\,\partial_\tau \bigg] \xi(\vx_1-\vx_2;\tau) 
\vs
& + \< X(\vx_3) \d(\vx_1) \d(\vx_2) \>\,,
\ea
where $\xi_{X Y}(r)$ denotes the cross-correlation between $X$ and $Y = a_{ij},\,\D\tau$.  
This expression gives the squeezed-limit three-point function of $X'$ and $\d'$ (in the \emph{primed} coordinates) in terms of derivatives of $\xi(\vr,\tau)$ and the three-point function of $X$ and $\d$ in the \emph{unprimed} coordinate frame.  As shown in \refapp{sqlimit}, we can derive an analogous expression in Fourier space, corresponding to the squeezed limit of the bispectrum:
\ba
B_{X'\d'\d'}(\vk_L,\vk_1,&\,\vk_2) \stackrel{\rm squeezed}{=} \vs
& \Bigg[P_{X a_{ij}}(k_L) \left(\d_{ij} + \frac{k_S^i k_S^j}{k_S^2} \frac{d}{d\ln k_S} \right) P(k_S,\tau) \vs
&\quad + P_{X\D\tau}(k_L) \frac{\partial}{\partial\tau} P(k_S,\tau) \Bigg]_{\vk_S = \vk_1+\vk_L/2} \vs
& + B_{X\d\d}(\vk_L,\vk_1,\vk_2)\,.
\ea
As before, $P_{XY}(k)$ denote cross-power spectra between $X$ and $Y$, while $P(k,\tau)$ denotes the power spectrum of $\rho$.  We can now further decompose $a_{ij}$ as
\be
a_{ij} = \frac1{n_D} a \d_{ij} + a^T_{ij},
\ee
where $n_D$ is the dimensionality of the space in which we define the
correlations, and $a_{ij}^T$ is traceless.  Allowing $n_D \neq 3$, in particular $n_D=2$ will become useful when dealing with projected observables such as the CMB.  We then obtain
\ba
B_{X'\d'\d'}(\vk_L,\vk_1,&\,\vk_2) \stackrel{\rm squeezed}{=} \vs
& \Bigg[ P_{X a}(k_L) P(k_S,\tau) \frac{d\ln (k_S^{n_D} P(k_S,\tau))}{d\ln k_S} \vs
& \quad + P_{X a^T_{ij}}(k_L) \frac{k_S^i k_S^j}{k_S^2} \frac{dP(k_S,\tau)}{d\ln k_S} \vs
& \quad + P_{X\D\tau}(k_L) \frac{\partial}{\partial\tau} P(k_S,\tau) \bigg]_{\vk_S = \vk_1+\vk_L/2} \vs
& + B_{X\d\d}(\vk_L,\vk_1,\vk_2) \,.
\label{eq:Bsqfinald2}
\ea
Note that the contribution of the trace of $a_{ij}$ scales as the logarithmic derivative of $k^{n_D} P(k)$, while the trace-free component $a^T_{ij}$ couples to the logarithmic derivative of $P(k)$ itself.  Instead of writing the bispectrum in terms of $k_S$, one can also expand in $k_L/k_1$, yielding zeroth order terms (with $k_S\to k_1$) and first-order terms $\propto k_L/k_1$.  The corresponding expression is given in \refapp{sqlimit}.  

The results of this section are useful when, say, $B_{X\d\d}$ is easy to calculate, while $B_{X'\d'\d'}$ is more readily related to observations.  This will be the case we encounter below.

\section{Conformal Fermi Normal Coordinates}
\label{sec:fnc}

Our goal is to derive what the observable consequences are (as opposed to coordinate artifacts) of a given primordial bispectrum. For this purpose we follow the steps in \reffig{overview}, starting with conformal Fermi Normal Coordinates (\fncb). As we discussed in the introduction, if we want to avoid studying second order cosmological perturbation theory, it is not possible to directly connect the statistics of perturbations in the standard FNC frame of galaxies or the CMB (at the time when the photons we observe were emitted) with the statistics calculated when all perturbations are superhorizon.  The reason is that the perturbative corrections to the Fermi frame metric are of order $(H x_F)^2$, where $x_F$ are the physical FNC coordinates, so that the metric in FNC is not approximately flat on the scales of superhorizon perturbations.  On the other hand, if we were to move from the global coordinates to FNC at a later time when the short wavelength modes have long re-entered the horizon, we would have to take into account their evolution after horizon entry at second order in the global coordinates. 

\begin{figure*}[t]
\centering
\includegraphics[width=\textwidth]{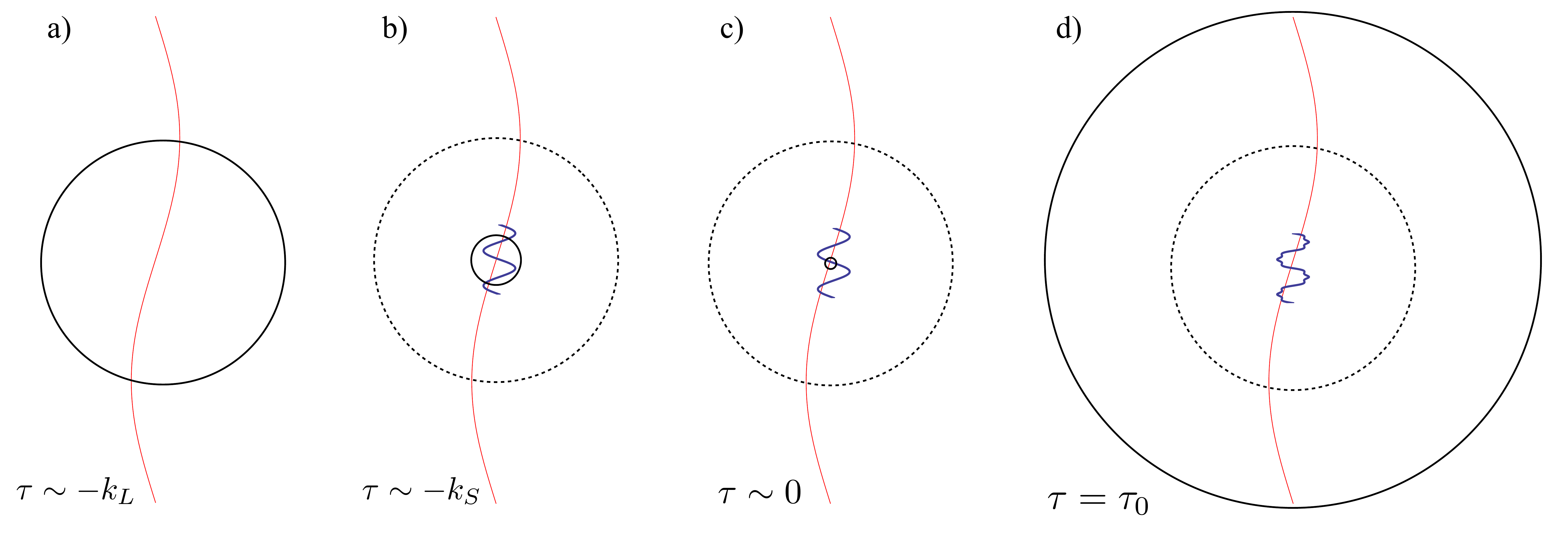}
\caption{Illustration of the \fncb~patch throughout cosmic history, in comoving units.  The wavy lines indicate perturbations.  The solid circles denote the comoving horizon
$1/aH$, which coincides with the size of the usual Fermi coordinate patch during inflation.  The dashed circles denote the size of the \fncb~patch, i.e.~the region within which the metric is of the form $\bar g_{\mu\nu}^F = a^2 \eta_{\mu\nu}$ with small corrections.
a) During inflation,
when the long wavelength perturbation $h_{ij}(\vk_L)$ is generated
(i.e.~leaves the horizon), thin red line.  
b) Later on during inflation; $h_{ij}(\vk_L)$ is far outside the horizon when the
short wavelength modes $\vartheta(\vk)$ are generated (thick blue line).  The
perturbations are generated well within the \fncb~patch corresponding to the 
long-wavelength mode.  
c) Near the end of inflation.  All perturbations are far outside the horizon.  
Nevertheless, the small-scale modes are still well within the \fncb~patch.
d) At observation time, after the long-wavelength mode has reentered the horizon 
(solid circle indicating present horizon).  The \fncb~patch now coincides with
the usual Fermi coordinate patch, which is much larger than the small-scale modes
which have been processed by nonlinear evolution since horizon entry (distorted
thick wavy line).
\label{fig:sketch}}
\end{figure*}

To overcome this obstacle, in this section we introduce a new coordinate frame which is valid throughout inflation
and the hot big bang era (assuming nothing dramatic happens to superhorizon perturbations during this time), and further connects easily with the FNC frame at late times. This provides a well-defined framework for
studying observables related to primordial non-Gaussianities in the squeezed limit. To visualize the advantage of \fncb, consider \reffig{sketch}. Each one of the four panels shows the Hubble scale (solid circle) and the region of validity of the \fncb~(dashed line) at a different moment in time. The long (red line) and short (blue line) modes are also shown for comparison. One can immediately appreciate the fact that from very early time when the long mode is still inside (or about to leave) the horizon all the way until late time when the long mode has re-entered the horizon, the region of validity of \fncb~amply encompasses the short modes. This is in contrast with standard \fnc~for which the presence of the Hubble scale makes this impossible. The primordial correlators, typically computed in comoving coordinates, take a simple form once both the short and long modes have left the horizon and cease to evolve. Any time after that moment and before the small-scale modes reenter the horizon constitutes a good time to transform from global conformal coordinates to \fncb.

Let us assume we are given the primordial correlations during inflation
in some gauge, where the metric is written as
\ba
ds^2 =\:& a^2(\tau)\left[\eta_{\mu\nu} + h_{\mu\nu} \right] dx^\mu dx^\nu
= a^2(\tau) \bar g_{\mu\nu} dx^\mu dx^\nu\,.
\label{eq:metric}
\ea
Here we have denoted the conformal metric with a bar.  In the
absence of perturbations, $h_{\mu\nu}=0$ and $\bar g_{\mu\nu} = \eta_{\mu\nu}$.  We will work up to linear order in metric perturbations.   

As in the usual Fermi normal coordinate construction, we consider a central timelike geodesic of a comoving observer.  However, instead of constructing the FNC 
with respect to $g_{\mu\nu}$, we construct \emph{conformal Fermi normal coordinates} (\fncb) with respect to $\bar g_{\mu\nu}$.  That is, within a region around the central geodesic, the metric in these coordinates is approximately 
$\bar g_{\mu\nu}^F = a^2(\tau_F) \eta_{\mu\nu}$.  The corrections to this metric, which determine the size of this region which we will call ``\fncb~patch'', are given by second derivatives of the metric perturbations.  Correspondingly, the size of the \fncb~patch is linked to the wavelength of the metric perturbations considered.  In the following, we will be interested in (long-wavelength) perturbations with wavenumber $k_L$, so that the size of the \fncb~patch is of order $k_L^{-1}$.

The transformation from coordinates $x^\mu$ to \fncb, for a patch centered at the spatial origin at time $x^0 = \tau_F$ is given by (see \refapp{FNCgen})
\ba
\bar x_F^0(x^\alpha) =\:& x^0 - \frac12 \int_0^{x^0} h_{00}(\tau) d\tau 
- (v_i + h_{0i}) x^i \vs
& - \frac14 \left[h_{0i,j} + h_{0j,i} + h'_{ij} \right] x^i x^j + \O[(x^i)^3] \\
\bar x_F^k(x^\alpha) =\:& x^k - v^k (x^0 - \tau_F)
+ \frac12 h^k_{\  i} x^i \vs
& + \frac14 \left[h^k_{\  i,j} + h^k_{\  j,i} - h_{ij}^{\  \  ,k}
\right] x^i  x^j + \O[(x^i)^3]\,.
\label{eq:FNCbtrans}
\ea
The inverse of this transformation is at linear order
\ba
x^0(\bar x^\alpha_F) =\:& \bar x^0_F + \frac12 \int_0^{\bar x^0_F} h_{00}(\tau) d\tau 
+ (v_i + h_{0i}) \bar x_F^i \vs
& - \frac14 \left[h^0_{\  i,j} + h^0_{\  j,i} - h'_{ij} \right] \bar x_F^i \bar x_F^j + \O[(\bar x_F^i)^3] \vs
x^k(\bar x^\alpha_F) =\:& v^k (\bar x_F^0 - \tau_F) + \bar x_F^k - \frac12 h^k_{\  i} \bar x_F^i \label{eq:FNCbtransI}\\
& - \frac14 \left[h^k_{\  i,j} + h^k_{\  j,i} - h_{ij}^{\  \  ,k}
\right] \bar x_F^i \bar x_F^j + \O[(\bar x_F^i)^3]\,.
\nonumber
\ea
The metric in \fncb~then becomes 
\ba
\bar g_{00}^F =\:& -1 + \frac12\left[ \partial_m\partial_l h_{00} + h_{lm}''
- 2 \partial_{(l} h'_{0m)} 
\right] \bar x_F^l \bar x_F^m 
\label{eq:metricFNCb}\\
\bar g_{0i}^F =\:& -\frac23 \left[ \partial_{[i} \partial_l h_{0m]}
+ \partial_{[m} h'_{li]} 
\right]
\bar x_F^l \bar x_F^m \vs
\bar g_{ij}^F =\:& \d_{ij} - \frac13 \left[
\partial_{[j} \partial_l h_{im]} + \partial_{[m} \partial_i h_{lj]}
\right]
\bar x_F^l \bar x_F^m.\nonumber
\ea
Here, $X_{(ab)} = (X_{ab}+X_{ba})/2$, $X_{[ab]} = (X_{ab}-X_{ba})/2$, and primes denote derivatives with respect to $\tau$, while spatial derivatives are with respect to $x^i$.  The linear and quadratic coefficients in the coordinate transform
\refeq{FNCbtrans} are uniquely determined by the requirement that the lowest
order contribution to $\bar g_{\mu\nu}^F - \eta_{\mu\nu}$ is order $(\bar x_F^i)^2$.  However, given that the quadratic corrections are determined by the
cubic order terms in the coordinate transformation, one might wonder whether 
the quadratic corrections to $\bar g_{\mu\nu}^F$ are unique.  One can show
(see \refapp{FNCU}) that $\bar g_{00}^F$ and $\bar g_{0i}^F$ are indeed unique.  
However, there is freedom to change the quadratic term in $\bar g_{ij}^F$
if one allows the spatial coordinate lines to be non-geodesic at order $(x^i)^3$.

In summary,
\be
\bar g_{\mu\nu}^F = \eta_{\mu\nu} + 
\O\left([h''_{ij}, \partial_j h'_{\mu i}, \partial_i \partial_j h_{\mu\nu}] \bar x_F^2\right).
\ee
Let us disregard the corrections $\propto h''_{ij},\:\partial_j h'_{\mu i}$ for the moment.  
The remaining corrections to the conformal
Fermi frame metric scale as second derivatives of the metric perturbation
multiplied by the spatial coordinate squared.  Thus, instead
of being order $(H x_F)^2$ as the usual FNC corrections , they are of 
order $(k_L \bar x_F)^2$,
where $k_L$ is the comoving wavenumber of the perturbation (in slight abuse
of notation, we drop the bar over $k$ since we will only be dealing with
comoving Fourier wavenumbers).  Within the \fncb~patch of size $\sim k_L^{-1}$, we can remove the
effects of the long-mode perturbation (in comoving coordinates) \emph{at all times} --- through horizon
exit and reentry.  

This very useful result only holds if $h''_{ij},\:\partial_j h'_{\mu i}$ are of order $\partial_i \partial_j h_{\mu\nu}$ or smaller, which is the case for all models of single field inflation in which the background is an attractor solution. In principle one can engineer a model in which, for a short period of time, the background is not close to an attractor solution, but rather it is evolving towards one. In this case, the superhorizon modes can have, for a short period of time, sizable time derivatives \cite{Namjoo:2012aa,Chen:2013aj}. For example, if $h''_{ij}$ is larger than $\partial_i \partial_j h_{\mu\nu}$, corresponding to significant superhorizon evolution of the spatial metric perturbation, then we cannot follow the conformal Fermi frame through the entire duration of inflation.  In case the evolving part of $h_{ij}$ is isotropic (or can be made so by a suitable
change of coordinates),
\be
h_{ij}(\v{0},\tau) = f(\tau) \d_{ij} + \O(\partial_i \partial_j h_{\mu\nu}),
\ee
where $\v{0}$ denotes the origin in comoving coordinates around which we
construct the \fncb~frame, then we can absorb its effect into a modified scale 
factor,
\be
\tilde a(\tau) = a(\tau) \left[1 - \frac12 f(\tau)\right]\,.
\ee
In a modified \fncb~frame constructed with the scale factor $\tilde a(\tau)$
the rapidly evolving part now disappears.  We thus see that in this case, 
a long-wavelength metric perturbation does not simply shift the time and rescale spatial coordinates, but rather corresponds to a change in the entire background cosmology.  This is possible only because the original ``unperturbed'' background cosmology, namely $f(\tau)=0$, was unstable (i.e.~not an attractor).  Given that the generation of small-scale perturbations depends on the background cosmology, we in general expect a non-trivial coupling between 
very long wavelength metric perturbations and small-scale modes in this case.  
If $h''_{ij}$ is \emph{not} isotropic, then we cannot absorb its effect
into a modified scale factor.  Instead, the different spatial coordinates
are rescaled differently, leading to an anisotropically expanding Universe
of Bianchi~I type.

Similar arguments apply to the case where $\partial_j h'_{\mu i}$ is larger than $\partial_i \partial_j h_{\mu\nu}$.  Here the lowest order effect will be a Bianchi~I-type Universe, since isotropy is necessarily violated.

\subsection{Bispectra in global and \fncb~coordinates}
\label{sec:bis}

Let us assume that we can calculate the statistics of a scalar field $\t(\tau,\vx)$
in the global coordinates \refeq{metric}.  Specifically, we consider 
fluctuations at a given epoch during inflation, typically evaluated right
after horizon crossing.  For simplicity, we will restrict
to a scalar field, although the extension to other spins is straightforward.  
Given the results from \refsec{sqlimit} and the coordinate transform \refeqs{FNCbtrans}{FNCbtransI}, we can then
immediately derive the transformation of the two-point correlation $\xi_\t(\vr,\tau)$ measured
in a patch around position $\vx_0$ at conformal time $\tau$ from the global
coordinates to the \fncb~frame.  Note that if $\t$ is non-Gaussian, then
the correlation function measured in a given patch will correlate with
long-wavelength perturbations.  

In the present case, the primed coordinate system of \refsec{sqlimit} is
$\bar x_F^\alpha$, while $x^\nu$ denote the global coordinates in the gauge chosen.  We can then read off $a^i_{\  j}$ and $\D\tau$ from \refeq{FNCbtransI}:
\ba
a^i_{\  j} =\:& \frac12 h^i_{\  j}
\label{eq:A_FNCb} \\
\D\tau =\:& \frac12 \int_0^{x'^0} h_{00}(\tau) d\tau\,.
\label{eq:Dtau_FNCb}
\ea
With this, \refeq{xitranslin} yields
\ba
\bar\xi(\bar\vr;\bar\tau) =\:& \label{eq:xitrans2}\\
\bigg[1\, - \:&\frac12 h_{ij} \bar r^i \partial_{\bar r}^j + \frac12 \left(\int_0^{\tau'} h_{00}d\tau\right)\partial_{\bar\tau} \bigg] \xi(\bar\vr, \bar\tau) \,,
\nonumber
\ea
where $\bar\xi(\bar\vr,\bar\tau)$ denotes the correlation function of $\t$ in the local \fncb~frame.  
Now consider the bispectrum $\<X(\vk_L) \t(\vk_1) \t(\vk_2)\>$, where $X$ is any perturbation.  The results from \refsec{sqlimit} immediately show how this bispectrum transforms into \fncb.  Denoting \fncb~quantities as $\bar X,\,\bar\t$, we obtain
\ba
B_{\bar X\bar\t\bar\t}(\vk_L,\vk_1,\vk_2;&\,\bar\tau) \stackrel{\rm squeezed}{=} \vs
& \Bigg[\frac12 P_{X h}(k_L) P_\t(k_S) \frac{d\ln (k_S^3 P_\t(k_S))}{d\ln k_S} \vs
& \quad + \frac12 P_{X h^T_{ij}}(k_L) \frac{k_S^i k_S^j}{k_S^2} \frac{dP_\t(k_S)}{d\ln k_S} \vs
& \quad + P_{X\D\tau}(k_L) \frac{\partial}{\partial\tau} P_\t(k_S) \bigg]_{\vk_S = \vk_1+\vk_L/2} \vs
& + B_{X\t\t}(\vk_L,\vk_1,\vk_2;\bar\tau) \,,
\label{eq:BXdd}
\ea
where all correlations on the r.h.s. are evaluated at $\bar\tau$.  Here $h = h^i_{\  i}/3$ is the trace of the spatial metric perturbation while $h^T_{ij}$ is the trace-free part.  The significance of this result will become clear in the application to single-field inflation which we will consider next.

\subsection{Single-field inflation in comoving gauge}
\label{sec:comgauge}

In the following, we restrict $h_{\mu\nu}$ to comoving gauge.  In the
notation of \cite{maldacena:2003}, we have
\ba
h_{00} =\:& -2 N_1 \vs
h_{0i} =\:& N_i \vs
h_{ij} =\:& 2\zeta \d_{ij} + h^T_{ij}(\tau,\vx),
\label{eq:hdecomp}
\ea
where $h^T_{ij}$ is tranverse-traceless and contains the tensor perturbations.  
For the attractor solution of single-field inflation, the constraint equations
in this gauge yield \cite{maldacena:2003}
\be
N_1 = \frac{\dot\zeta}{H} = \frac{\zeta'}{aH} \propto \frac{k^2}{(aH)^2}\,.
\label{eq:N1}
\ee
To order $k_L^2/ k_S^2$, we can thus neglect the contribution from the time shift
$\D\tau = - \int N_1 d\tau$.  Note that this is merely a consequence of the particular gauge chosen.  The only remaining contribution to the transformation of the bispectrum [\refeq{BXdd}] then comes from $\zeta$ and the tensors $h^T_{ij}$.  For the scalar contribution, which we indicate with and ``S'' over the equal sign, we obtain
\ba
B_{\bar X\bar\t\bar\t}(\vk_L,\vk_1,\vk_2) \stackrel{\rm S}{=}\:& P_{X \zeta}(k_L)  P_\t(k_S) \left[3 +  \frac{\partial\ln P_\t(k_S)}{\partial\ln k_S}\right] \vs
& + B_{X\t\t}(\vk_L,\vk_1,\vk_2), 
\ea
where throughout $\vk_S = \vk_1 + \vk_L/2 = (\vk_1-\vk_2)/2$.  We are
in particular interested in the bispectrum of the curvature perturbation
$\zeta$, i.e.~$\t = \zeta$.  Since $\zeta$ is not a scalar, our
derivation in \refsec{sqlimit} does not strictly apply.  However, it
is straightforward to show that for the purposes of this transformation,
the small-scale $\zeta$ modes behave as a scalar.  Recall that only
the spatial transformation is relevant at the order we are interested.  
The comoving gauge condition, $\d\phi = 0$ where $\d\phi$ is the inflaton
perturbation is thus still satisfied in the \fncb~frame.  Further,
the spatial components of the metric transform as
\be
\bar g_{ij}^F(\bar x_F) = \frac{\partial x^k}{\partial\bar x_F^i}
\frac{\partial x^l}{\partial\bar x_F^j} g_{kl}(x)\,.
\ee
We now write $\zeta(x) = \zeta_L(x) + \zeta_S(x)$, separating into 
long- and short-wavelength pieces on the scale of the patch within which
the correlation function is measured.  Then, the transformation to \fncb~removes $\zeta_L$ up to second derivatives, while $\zeta_S$ is not affected
since it does not contain any long-wavelength components.  We thus obtain
\ba
\bar \zeta(\bar x_F) =\:& \zeta_S(x) + \O(\partial_i\partial_j \zeta_L) \vs
\bar\zeta_S(\bar x_F) =\:& \zeta_S(x)\,.
\label{eq:zetatrans}
\ea
Thus, the short-wavelength $\zeta$ perturbations transform effectively
as a scalar [\refeq{dtransf}].  We obtain for the bispectrum of curvature perturbations in the \fncb~frame:
\ba
B_{\bar\zeta\bar\zeta\bar\zeta}(\vk_L,\vk_1,\vk_2) =\:& P_\zeta(k_L) P_\zeta(k_S) \frac{\partial\ln (k_S^3 P_\zeta(k_S))}{\partial\ln k_S} \vs
& + B_{\zeta\zeta\zeta}(\vk_L,\vk_1,\vk_2),
\label{eq:Bzetatrans}
\ea
where $B_{\zeta\zeta\zeta}$ is the three-point function calculated in comoving gauge.  As shown in \cite{maldacena:2003,CreminelliEtal11}, this is in the squeezed limit given by
\be
B_{\zeta\zeta\zeta}(\vk_L,\vk_1,\vk_2) = - [n_s-1] P_\zeta(k_L) P_\zeta(k_S)\,,
\label{eq:consrel}
\ee
where $n_s$ is defined through $P_\zeta(k) \propto k^{-4+n_s}$.  \refeq{consrel} is usually referred to as the \emph{consistency relation}.  We now see that the first term in \refeq{Bzetatrans} exactly cancels the contribution from $B_{\zeta\zeta\zeta}$, leading to 
\be
B_{\bar\zeta\bar\zeta\bar\zeta}(\vk_L,\vk_1,\vk_2) = \O\left(\frac{k_L}{k_S}\right)^2\,.
\ee
We thus conclude that \emph{in single-field inflation, the bispectrum in the squeezed limit is zero in the conformal Fermi frame, with corrections going as $(k_L/k_S)^2$.}  

The case for non-scalar metric perturbations follows analogously.  
Decomposing the long-wavelength metric perturbation into polarization states,
\be
h^{T}_{ij}(\vk_L) = \sum_{s=+,\times} e^s_{ij}(\hat\vk_L) h^T_s(\vk_L),
\ee
we obtain (the ``T'' over the equal sign now stands for tensor)
\ba
B_{\bar X \bar\t\bar\t}&(\vk_L,\vk_1,\vk_2) \stackrel{\rm T}{=} \vs
& \frac12 \sum_s P_{X h_s}(k_L) P_\t(k_S) e^s_{ij}(\hat\vk_L) \hat k_S^i \hat k_S^j  \frac{\partial\ln P_\t(k_S)}{\partial\ln k_S} \vs
& + B_{X\t\t}(\vk_L,\vk_1,\vk_2).
\ea
Assuming that the different polarization states are statistically independent,
we obtain for the tensor-scalar-scalar bispectrum [again using the transformation property \refeq{zetatrans}]
\ba
& B_{\bar h_s \bar\zeta\bar\zeta}(\vk_L,\vk_1,\vk_2) = \vs
&\quad \frac12 P_{h_s}(k_L) P_\zeta(k_S) e^s_{ij}(\hat\vk_L) \hat k_S^i \hat k_S^j  \frac{\partial\ln P_\zeta(k_S)}{\partial\ln k_S} \vs
&\quad\quad + B_{h_s\zeta\zeta}(\vk_L,\vk_1,\vk_2).
\label{eq:Bhzztrans}
\ea
The squeezed-limit bispectrum in comoving gauge was also derived in \cite{maldacena:2003}:
\ba
& B_{h_s\zeta\zeta}(\vk_L,\vk_1,\vk_2) = - P_{h_s}(k_L) e^s_{ij}(\hat\vk_L) k_S^i k_S^j \frac{d}{dk_S^2} P_\zeta(k_S) \vs
& \quad = - P_{h_s}(k_L) e^s_{ij}(\hat\vk_L) \hat k_S^i \hat k_S^j \frac12 \frac{d}{d\ln k_S} P_\zeta(k_S)\,.
\label{eq:consrelt}
\ea
We again see that the two terms in \refeq{Bhzztrans} cancel in single-field inflationary models.  Thus,
\emph{in single-field inflation, the tensor-scalar-scalar bispectrum vanishes
in the squeezed limit in \fncb, so that there are no
correlations between long-wavelength tensor modes and small-scale fluctuations
in this frame.}  The lowest correction are again of order $(k_L/k_S)^2$.  

We can phrase the main result of this section as follows: at leading order,
the squeezed-limit three-point correlations in single-field inflation,
which obey the ``consistency relation'', are equivalent to the statement
that there is no correlation between infinitely long and short wavelength 
modes in the conformal Fermi frame, specifically,
\be
B_{\bar h\bar\zeta\bar\zeta}(\vk_L,\vk_1,\vk_2) = \O\left(\frac{k_L}{k_S}\right)^2,
\label{eq:3ptschem}
\ee  
where $h$ stands for any component of $h_{ij}$, and $k_S = |\vk_1-\vk_2|/2$.  

Since this latter is a physical, gauge-invariant statement, it is expected to hold not only at leading order in spacetime perturbations but at higher orders as well.  The FNC approach can also help elucidate why there is no such correlation up to order $k_L^2$ in single field models.  One might wonder why such a correlation cannot be imprinted at early times when the long wavelength perturbation was inside the horizon, $k_L \gg a H$.  Far inside the horizon, we can
neglect gravity and are essentially dealing with a scalar field in vacuum
which adiabatically tracks the slowly evolving background, and a given mode is only excited
once its wavelength becomes of order the horizon.  Contributions to correlations
from within the horizon are exponentially suppressed \cite{Senatore:2012ya,Senatore:2012wy} (see also \cite{flauger/green/porto}).  On the other hand,
when $k_S \sim a H$, the long-wavelength mode is far outside the horizon and
its effects can be removed by a coordinate transformation up to order $k_L^2$.

\section{Connection to late-time observations}
\label{sec:latetime}

As explained in \refsec{fnc}, we can follow the \fncb~patch all the way
through the end of inflation and horizon re-entry of both long and short
modes, provided that perturbations do not evolve significantly when 
they are outside the horizon.  In this section, we show how the
squeezed-limit correlations in the \fncb~frame can be related to observations
made from Earth today.

Let us assume we observe correlations of some field $\d(\vx)$
at late times, i.e.~at or after recombination.  In linear theory, assuming adiabatic initial conditions, we can relate $\d$ to the curvature perturbation through some transfer function $\alpha$,
\be
\d(\vk,\tau) = \alpha(k,\tau) \zeta(\vk).
\label{eq:deltazeta}
\ee
For matter density perturbations, $\alpha$ is defined in \refeq{alpha} below.  
Similarly, we assume the long-wavelength perturbation is evolved
with some transfer function,
\be
h_{ij}(\vk_L, \tau) = G(\tau) h_{ij}(\vk_L),
\label{eq:zetatransfer}
\ee
where we have assumed the $k_L \to 0$ limit (the lowest order $k_L$ dependence
of $G(\tau)$ will be $k_L^2$; of course, $G$ will differ for scalar and tensor perturbations).  
As long as superhorizon perturbations evolve slowly (in the sense that
$h'', \partial_j h'$ are smaller than or of order $\partial_i \partial_j h$), 
the conformal Fermi coordinate patch, i.e.~the
region over which corrections to the \fncb~metric are small,
is essentially constant in (comoving) size throughout horizon exit and reentry of the
short-wavelength modes.  As we have seen, this applies in particular to single-field
inflation models.  Thus, at the conformal time $\tau_{\rm em}$ at which the photons we observe
today were emitted, the metric in \fncb~is 
\be
\bar g^F_{\mu\nu} = a^2(\tau_{\rm em}) \eta_{\mu\nu} + \O(\partial_i \partial_j h)\,.
\ee
If we transform coordinates through
\ba
x_F^0 =\:& \int^{\tau} a(\tau')d\tau' \equiv t(\tau) \vs
x_F^i =\:& a(\tau) \bar x_F^i,
\label{eq:fncbtofnc}
\ea
the metric in the coordinates $\{ x_F^\mu \}$ becomes 
\be
g^F_{\mu\nu} = \eta_{\mu\nu} + \O[(H x_F^i)^2, (\partial\partial h_{\mu\nu}) x_F^2].
\ee
In other words, $x_F^\mu$ are the usual Fermi coordinates (FNC) defined around
the same timelike geodesic as the \fncb.  Depending on whether the long-wavelength modes for which we constructed the \fncb~patch have entered the horizon, either the order $H$ or the order $\partial\partial h_{\mu\nu}$ corrections will be dominant; however, this is not relevant for the discussion that follows.  
\refeq{fncbtofnc} corresponds to a rescaling of the time coordinate, leaving the timelike unit vector
and $\tau = $~const hypersurfaces unchanged, and a time-dependent rescaling of
the spatial coordinates.  Since the spatial rescaling
is the same everywhere on a $x_F^0 = $~const hypersurface, this implies 
that, for models that obey the consistency condition, \refeq{3ptschem} is still valid at late times in the FNC frame:
\be
B_{h_F \d_F\d_F}(\vk_L,\vk_1,\vk_2; \tau_{\rm em}) = \O\left(\frac{k_L}{k_S}\right)^2\,,
\label{eq:3ptschem_e}
\ee
where $h_F,\,\d_F$ denote perturbations in the FNC frame.  
Recall that the squeezed limit of three-point functions corresponds to the modulation of local two-point functions by long-wavelength perturbations.  
Thus, another way of phrasing this result is that a surface of constant correlation
in the \fnc~frame, $\xi_{\d_F}(r_F) = $~const, defines a standard ruler---a fixed spatial scale---as considered in \cite{stdruler}, with corrections proportional to second
derivatives of $h_{\mu\nu}$ only.  Hence, in standard single-field 
inflation and any other case where \refeq{3ptschem} holds, the ruler scale 
$r_F$ is statistically the same everywhere on a $x^0_F=$~const hypersurface, 
and there is no correlation with long-wavelength perturbations.  

The \emph{apparent} correlations induced between long-wavelength modes
and small-scale correlations are then given by the ruler perturbations
derived in \cite{stdruler}.  We can use their results together with 
\refsec{sqlimit} to derive these
contributions to observed squeezed-limit three-point functions, and
make the connection with known results.  

In general, there are two effects modifying the observed two-point correlation within a given patch.  
First, there is the transformation from the local FNC $x_F^\mu$ to the 
observed comoving coordinates $\tilde x^\mu$, which are inferred from the
observed position $\vnhat$ of the source in the sky and its redshift
$\tilde z$ through $\tilde x^0 = \tau_0 - \bar\chi(\tilde z)$, $\tilde x^i = \nhat^i \bar\chi(\tilde z)$, where $\bar\chi$ is the
comoving distance-redshift relation in the background (in case of the CMB, this is slightly modified, as we will discuss in \refsec{CMBbis}).  Let us assume we observe a scalar field $\tilde\o(\tilde x)$ (whose perturbation is $\d$).  Then, as described in \refsec{sqlimit}, $\tilde\o$ is given in terms of the field in the Fermi frame $\o$ as
\be
\tilde\o(\tilde x) = \o(x_F(\tilde x))\,.
\ee
At fixed observed redshift, we can write the transformation from $x_F$ to $\tilde x$ as
\ba
\tilde x^0 =\:& x_F^0 + \frac1{\tilde a \tilde H} \T \vs
\tilde x^i =\:& \frac1{\tilde a} A^i_{\  j} x_F^j\,.
\ea
Here $\tilde a = (1+\tilde z)^{-1}$, $\T$ is the perturbation in proper time from a constant observed redshift surface \cite{T}, and $A^i_{\  j}$ can be seen 
as the generalization to three dimensions of the magnification matrix.  Note that $\T$ and $A^i_{\  j}$
are gauge-invariant.  Specifically, using the notation of \cite{stdruler}, $A_{ij}$ is given by
\ba
A_{ij} =\:& \d_{ij} + a_{ij} \vs
a_{ij} =\:& \C \nhat_i \nhat_j + \nhat_{(i} \P_{j)k} \B^k + \P_{ik} \P_{jl} \A^{kl},\label{eq:rulerpert}
\ea
where $\P^{ij} = \d^{ij} - \nhat^i\nhat^j$ is the projection operator perpendicular
to the line of sight.  $\C,\,\B_i,$ and $\A_{ij}$ are the gauge-invariant
ruler perturbations derived in terms of the metric perturbations in 
\cite{stdruler}.  For example, the transverse matrix $\A_{ij}$ contains
the magnification and shear.  

The second effect is a rescaling of the observed field by projection
effects.    If there are
multiplicative projection effects, then we have
\be
\o_{\rm obs}(\tilde x) = \left[1 + c(\tilde x)\right] \tilde\o(\tilde x)
= \left[1 + c(\tilde x)\right] \o(x_F(\tilde x))\,.
\label{eq:ctransf}
\ee 
Here we think of $c$ as averaged over the patch within which we measure
the correlation function of small-scale perturbations.  
For example, in case $\o$ is the number density of some tracer, then 
projection effects such as gravitational lensing modify the physical
volume that corresponds to a fixed region in the observed coordinates
$\tilde x$, thus rescaling the number density.  Gravitational redshift
and ISW effect lead to an analogous rescaling in case of the CMB.  
The factor $1+c$ also rescales the fluctuations in $\o$ within the region
considered, and correspondingly rescales the correlation function by $(1+c)^2$, leading to
\ba
 \tilde\xi(\tilde \vr, \tilde\tau) =\:& \left[1 + 2 c(\tilde x)\right] \xi(\vr_F, \tau_F) \vs
=\:& \left[1 - a_{ij}(\tilde x) \tilde r^i \partial_{\tilde r}^j + \frac1{\tilde a\tilde H} \T(\tilde x)\partial_\tau  + 2 c(\tilde x)\right] \xi(\tilde\vr;\tilde\tau)
\ea
where we have used \refeq{xitranslin} in the second line.  
Straightforward application of the results of \refsec{sqlimit} then 
yields the bispectrum of the observed density perturbation $\tilde\d$ in
the squeezed limit, in terms of the projection effects and  the bispectrum of $\d_F$ in the \fnc~frame:
\ba
B_{\tilde\d\tilde\d\tilde\d}(\vk_L, & \,\vk_1, \vk_2) = B_{\d_F\d_F\d_F}(\vk_L,\vk_1,\vk_2) \vs 
&  + 2 P_{\d_F c}(k_L) P_{\d_F}(k_S) + \frac1{\tilde a\tilde H} P_{\d_F \T}(k_L) \frac{\partial}{\partial\tau} P_{\d_F}(k_S) \vs
&  +\frac1{3} \frac{\partial \ln (k_S^3 P_{\d_F}(k_S))}{\partial\ln k_S} 
P_{\d_F a}(k_L) P_{\d_F}(k_S) \vs
&  + P_{\d_F a^T_{ij}}(k_L) \hat k_S^i \hat k_S^j \frac{\partial P_{\d_F}(k_S)}{\partial \ln k_S},
\label{eq:Bdsqueezed}
\ea
where $\vk_S = \vk_1 + \vk_L/2$.  The bispectrum of a two-dimensional projected field (in the flat-sky limit) correspondingly becomes
\ba
&B_{\tilde\d\tilde\d\tilde\d}(\vl_L, \vl_1, \vl_2) = B_{\d_F\d_F\d_F}(\vl_L,\vl_1,\vl_2) \vs 
& + 2 C_{\d_F c}(\l_L) C_{\d_F}(\l_S) 
+ \frac1{\tilde a\tilde H} C_{\d_F \T}(\l_L) \frac{\partial}{\partial\tau} C_{\d_F}(\l_S) \vs
& +\frac12 \frac{\partial \ln (\l_S^{2} P_{\d_F}(\l_S))}{\partial\ln \l_S} 
C_{\d_F a}(\l_L) C_{\d_F}(\l_S) \vs
& + C_{\d_F a^T_{ij}}(\l_L) \hat \l_S^i \hat \l_S^j \frac{\partial C_{\d_F}(\l_S)}{\partial \ln \l_S}\,,
\label{eq:Bdsqueezed_2d}
\ea
where again $\vl_S = \vl_1 + \vl_L/2$.

The bispectrum $B_{\d_F\d_F\d_F}$ in FNC frame is equivalent up to transfer
functions (and radial projection, in the 2D case) to the bispectrum in \fncb, 
which as we have seen vanishes in the
squeezed limit for single-field inflation.  In this case, the terms due
to ``projection effects'' are the only remaining contributions.  

In the following we present two applications of \refeqs{Bdsqueezed}{Bdsqueezed_2d}: the CMB bispectrum in the squeezed limit, and the scale-dependent non-Gaussian halo bias.

\subsection{CMB bispectrum}
\label{sec:CMBbis}

In this section, we illustrate the main result of the previous section
on the observed squeezed-limit bispectrum, \refeq{Bdsqueezed_2d}, with
the CMB.  We will make direct connection with the results of \cite{CreminelliPitrouVernizzi,Bartolo:2011wb}.  A more sophisticated and accurate treatment which is based on an closely related approach has been presented in \cite{Lewis:2012}.

We assume single-field inflation so that the \fncb~frame
contribution vanishes.  We will adopt the conformal-Newtonian gauge in this section,
\be
ds^2 = a^2(\tau)[ -(1+2\Phi) d\tau^2 + (1-2\Psi) d\vx^2 ]\,,
\label{eq:metriccN}
\ee
although it is straightforward to derive the results in a general gauge.

The observed CMB photons originate from the last scattering surface, which
occurred at a fixed physical age $t_*$ of the Universe, that is, 
at constant proper time $t_F = t_*$ for the comoving primordial plasma.  
The proper time of a comoving source passing through $\vx$ at coordinate time $\tau$ is given in the metric convention \refeq{metriccN} by
\ba
t_F|_{\tau,\vx} =\:& \bar t(\tau) + \int_0^{\tau} \Phi(\vx,\tau') a(\tau') d\tau' \,,
\label{eq:tF}
\ea
where $\bar t(\tau)$ and $\bar\tau(t)$ are the physical-time -- conformal time relations
in the background.  

The value of $t_*$ is obtained by combining atomic physics with the mean observed temperature of the CMB today.  The CMB temperature perturbations on scales that were 
super-horizon at recombination ($\l \lesssim 100$) originate entirely from projection effects;  in other words, the large-scale CMB temperature perturbations can be seen as a special case of the evolving ruler described in \cite{T}.  Essentially, the standard ruler is in this case given by the photon occupation number $I_\nu/\nu^3$.  In fact, since lensing conserves surface brightness, the only contribution to the fractional CMB temperature perturbation $\Theta(\vnhat)$ is the redshift perturbation on a fixed proper time surface, which is equivalent to minus the quantity $\T$ defined in \cite{T}:
\be
\Theta(\vnhat) \equiv \frac{T(\vnhat)}{\bar T} - 1 = \frac{d\ln T(a)}{d\ln a} \T = -\T \,
\label{eq:ThetaCMB1}
\ee
where have used that $T \propto a^{-1}$ for a free-streaming blackbody.  As shown in \cite{T}, this immediately yields the CMB temperature perturbation in the gauge \refeq{metriccN} as
\ba
\Theta(\vnhat) =\:& - \tilde H_* \int_0^{t_*} \Phi[\vx,\bar\tau(t')] dt' + H_0 \int_0^{t_0} \Phi(\v{0},\bar\tau(t)) dt \vs
& - \Phi_o + \Phi - v_\parallel + v_{\parallel o} + \d\nu_{\rm ISW} \vs
\simeq \frac13 \Phi -\:& v_\parallel + H_0 \int_0^{t_0} \Phi(\v{0},\bar\tau(t)) dt - \Phi_o + v_{\parallel o} + \d\nu_{\rm ISW}
\label{eq:ThetaCMBcN}\\
\d\nu_{\rm ISW} =\:&  \int_0^{\chis} d\chi\left[\Phi'+\Psi'\right]\,.
\ea
In the second line, we have assumed that the relevant perturbations are superhorizon, $\Phi(\vx) \approx$~const (since we are interested in $\l \lesssim 100$ where the acoustic contributions can be neglected), and that recombination happened long after matter-radiation equality so that $H_* t_* = 2/3$.

\subsubsection{Transforming from \fnc~to observer coordinates}

As we have seen, the CMB temperature on large scales is modified
by the temperature perturbation \refeq{ThetaCMBcN}, which is entirely
due to the effects on photons as they propagate from the last scattering
surface to the observer.  Thus, the CMB temperature that would be measured in the
local Fermi frame at emission is rescaled by $1+\Theta(\vnhat)$ in the observer frame, so that in the
notation of \refsec{latetime}, $c(\vnhat) = \Theta(\vnhat)$.  

The second ingredient is the coordinate transformation from \fnc~to 
the observer frame.  Projected onto the sky, the matrix $a_{ij}$ becomes the usual weak lensing distortion tensor $\A_{ij}$, which we can write to linear order as
\be
a_{ij} = \left(\begin{array}{cc}
\M/2 + \g_1 & \g_2 \\
\g_2 & \M/2 - \g_1
\end{array}\right).
\ee
In particular, $a = \M$.  The magnification (a gauge-invariant quantity) is easily adapted from the results of
\cite{stdruler}.  The key difference is the meaning of the perturbation to the logarithm of the scale factor at emission, which here we call $\D\ln a_*$.  By definition,
\be
\D\ln a_* \equiv \frac{a(x^0_{\rm em})}{a_*} - 1.
\ee  
Whereas the quantity $\D\ln a$ of \cite{stdruler} is derived for photons arriving with a fixed observed redshift, in the present case we need the corresponding expression for a constant proper time at emission.  This is easily derived from the expression for the proper time \refeq{tF}.  Requiring $t_F = t_*$ and solving for $\tau$ yields
\ba
a(x^0_{\rm em}) =\:&  a\left[\bar \tau\left(t_* - \int_0^{\tau_*} \Phi(\vx,\tau)\, a d\tau\right)\right] \vs
=\:& a_* \times\left[ 1 - H_* \int_0^{\tau_*} \Phi(\vx,\tau)\, a d\tau\right]\,,
\label{eq:aem}
\ea
which leads to
\be
\frac{a(x^0_{\rm em})}{a_*} = 1 - H_*\int_0^{\tau_*} \Phi(x,\tau) a d\tau\,,
\ee
so that
\be
\D\ln a_* = -H_*\int_0^{\tau_*} \Phi(x,\tau) a d\tau\,.
\ee
We now use this result in the general-gauge expression Eq.~(51) of \cite{stdruler},
\ba
\M =\:& - 2\D\ln a_* - \frac12 \left(h^i_{\  i} - h_\parallel\right) 
+ 2\hat\k - \frac{2}{\chis} \D x_\parallel\,.
\label{eq:mag}
\ea
In \refeq{mag}, $\hat\k$ is the coordinate convergence, i.e.~the transverse 
(with respect to the line of sight) divergence of the transverse displacements,
and $\D x_\parallel$ is the displacement along the line of sight.  In the metric \refeq{metriccN} we have \cite{stdruler} 
\ba
\hat\k =\:& - v_{\parallel o}
+ \frac12 \int_0^{\chis} d\chi\,\frac{\chi}{\chis} 
 (\chis-\chi)
\nabla_\perp^2 \left(\Phi + \Psi\right) 
\label{eq:kcN} \\
\D x_\parallel =\:& \int_0^{\chis} d\chi\left[ \Phi + \Psi \right]
- \frac{1}{a_* H_*} \D \ln a_* \vs
&- \int_0^{t_0} \Phi(\v{0},t) dt \label{eq:Dxpar}\\
\D\ln a_* =\:& - \frac23 \Phi\,,
\ea
where $H_* = H(a_*)$, and we have made the same assumptions about superhorizon perturbations and matter domination as in \refeq{ThetaCMBcN}.  The magnification then becomes
\ba
\M =\:& \frac43 \Phi  - (-2\Psi) + 2\hat\k - \frac{2}{\chis} \D x_\parallel \vs
=\:& \frac{10}3 \Phi + 2\hat\k - \frac{2}{\chis} \D x_\parallel \,.
\label{eq:MCMB}
\ea
Here we have used that $\Psi=\Phi$ during matter domination.  

Using the relation between convergence $\k$ and shear $\g$ 
in  $\l$-space, we then have in the flat-sky approximation
\ba
C_{T c}(\l) =\:& C_{TT}(\l) \vs
C_{T a}(\l) =\:& C_{T\M}(\l) = \frac{10}3 C_{T \Phi_{\rm em}}(\l) + 2 C_{T\hat\k}(\l) 
 - \frac2{\chis} C_{T \D x_\parallel}\vs
\approx\:& 10 C^{\rm no~ISW}_{TT}(\l) + 2 C_{T\hat\k}(\l) \label{eq:CTM}\\
C_{T h^T_{ij}}(\l) =\:& C_{T \hat\k}(\l) \left[2\frac{\l_i \l_j}{\l^2} -\d_{ij}\right].
\ea
In \refeq{CTM}, we have approximated $C_{T \Phi_{\rm em}}(\l)/3$ as the
CMB temperature power spectrum without ISW contribution, and neglected
the contribution from the line-of-sight displacement $\D x_\parallel$.  
The latter is small due to cancelation along the line of sight except
for the very smallest $\l$ for the time delay [first term in \refeq{Dxpar}],
and suppression by $(H_*\chis)^{-1} \sim \sqrt{1089}$ in case of the second
term of \refeq{Dxpar} (the third term in \refeq{Dxpar} only contributes to the monopole).  See \cite{Lewis:2012} for a quantitative evaluation of these contributions.  Finally, for the contribution from the time shift, $C_{T \T}(\l_L) = - C_{TT}(\l_L)$ since $\Theta = -\T$.  However, the CMB power spectrum observed today $C_{TT}(\l_S)$ only evolves on the Hubble scale today, so that 
\be
\frac1{\tilde a\tilde H} C_{T\T}(\l_L) \frac{\partial}{\partial\tau} C_{TT}(\l_S) \sim \frac{H_0}{\tilde a\tilde H} C_{TT}(\l_L) C_{TT}(\l_S)\,.
\ee
Thus, this contribution is suppressed by $\sim 1/\sqrt{1089}$ with respect
to the leading contributions, and we will neglect it in what follows.

\subsubsection{Squeezed-limit CMB bispectrum}

Inserting these results into \refeq{Bdsqueezed_2d} then yields
\ba
& B(\l_L, \l_1, \l_2) = 
 2 C_{T c}(\l_L) C_{TT}(\l_S) \vs
&\quad + C_{T a}(\l_L) C_{TT}(\l_S) \frac12\frac{\partial \ln \l_S^2 C_{TT}(\l_S)}{\partial\ln \l_S} \vs
&\quad + C_{T a^T_{ij}}(\l_L)  \hat \l_S^i \hat \l_S^j \frac{\partial C_{TT}(\l_S)}{\partial \ln \l_S} \vs
& = \Bigg\{2 C_{TT}(\l_L) + \left(5 C^{\rm no~ISW}_{TT}(\l_L) + C_{T\hat\k}(\l_L) \right)
\frac{\partial \ln \l_S^2 C_{TT}(\l_S)}{\partial\ln \l_S}
\vs
&\quad\quad  + \cos 2\theta\: C_{T\hat\k}(\l_L) \frac{\partial\ln C_{TT}(\l_S)}{\partial\ln \l_S} \Bigg\} C_{TT}(\l_S),
\ea
where $\vl_S = \vl_L + \vl_1/2$, and 
$\cos\theta = \hat{\vl}_L\cdot\hat{\vl}_S$.  Apart from the
very small contribution from $\D x_\parallel$, this is an exact result for the
squeezed-limit CMB bispectrum as long as the approximation of the
Sachs-Wolfe limit is accurate for modes of wavenumber $\l_L$ ($\l_L \lesssim 100$).  Note that no such assumption has been made about the short-wavelength modes $\l_S$.  
The cross-correlation between CMB temperature and $\hat\k$ can further
be decomposed as
\be
C_{T\hat\k}(\l_L) = \frac13 C_{\Phi_{\rm em}\hat\k}(\l_L) 
+ C_{\d\nu_{\rm ISW}\hat\k}(\l_L),
\ee
i.e.~an early-time correlation with the potential $\Phi_{\rm em}$, and
a late-time correlation between $\hat\k$ and the ISW contribution.  The
latter contribution has recently been detected by the Planck satellite \cite{PlanckISW}.  

In order to compare with \cite{CreminelliPitrouVernizzi,Bartolo:2011wb}, we neglect
the ISW-lensing correlation, and use
the result of \cite{BoubekeurEtal} for early-time correlations in the
Sachs-Wolfe regime,
\be
\frac13 C_{\Phi_{\rm em}\hat\k}(\l_L) = -6\: C^{\rm no~ISW}_{TT}(\l_L).
\label{eq:CTk}
\ee
Further, we neglect the distinction between $C_{TT}(\l_L)$ and 
$C_{TT}^{\rm no~ISW}(\l_L)$.  We then have
\ba
& B(\l_L, \l_1, \l_2) \vs
& = \bigg[2 C_{TT}(\l_L) - C_{TT}(\l_L) 
\frac{\partial \ln \l_S^2 C_{TT}(\l_S)}{\partial\ln \l_S}
\vs
&\quad - 6\cos 2\theta\: C_{TT}(\l_L) \frac{\partial\ln C_{TT}(\l_S)}{\partial\ln \l_S} \bigg] C_{TT}(\l_S)
\vs
& = C_{TT}(\l_L) C_{TT}(\l_S) \left(1 + 6 \cos2\theta\right)\left[2 - \frac{\partial \ln \l_S^2 C_{TT}(\l_S)}{\partial\ln \l_S} \right].
\ea
This agrees with the final result of \cite{CreminelliPitrouVernizzi}, \chg{while it differs from \cite{Bartolo:2011wb} because a subset of the lensing contributions was neglected there.} 

One can show (App.~C of \cite{stdruler}) that the contribution of metric
perturbations with wavenumber $k$ to the 
CMB temperature as well as the ruler perturbations $\M,\:\g$ scale
as $(k/H_0)^2$ in the limit $k\to 0$.  The contributions in the low-$k$ limit can be interpreted as the lowest order corrections to our local conformal Fermi patch.  Unless there is some physical coupling between currently superhorizon and subhorizon modes \cite{SchmidtHui}, the observable imprint of \emph{any} superhorizon perturbation (scalar or tensor) is suppressed by $(k/H_0)^2$.

\subsection{Non-Gaussian halo bias}
\label{sec:bias}

For Gaussian initial conditions, the distribution of large-scale structure
tracers (such as galaxies, clusters, etc.) in the large-scale limit follows 
the distribution of matter.  More accurately, this holds on a constant proper 
time slice \cite{BaldaufEtal,PBSpaper}.  While the abundance of tracers 
depends on the amplitude and shape of local small-scale fluctuations, in the
Gaussian case these are statistically the same everywhere.  That is, at fixed
proper time in their respective FNC~frame, all observers see statistically the
same small-scale fluctuations.  These then do not
contribute to correlations in the tracer abundance on large scales.  
Non-Gaussianity in the primordial perturbations can however couple small-scale 
fluctuations to large-scale perturbations.  As we discussed in \refsec{sqlimit}, 
this in fact precisely corresponds to the squeezed limit of (at lowest order)
the three-point function.  Thus, neglecting for the time being any projection
effects in going from the FNC~frame to the observed positions and redshifts of tracers, there is a non-Gaussian scale-dependent bias if and only if the 
amplitude of small-scale fluctuations in the FNC~frame correlates with long-wavelength perturbations---that is, if there is a non-zero squeezed-limit bispectrum $B_{\d_F\d_F\d_F}$, where $\d_F$ denotes the matter density perturbation in the FNC~frame.  

Using \refeqs{deltazeta}{zetatransfer}, the bispectrum of $\d_F$ is related to the bispectrum of curvature perturbations in the \fncb~frame through
\ba
B_{\d_F \d_F \d_F}(\vk_L,\vk_1,\vk_2;\tau_{\rm em}) =\:& \alpha(k_L,\tau_{\rm em}) \alpha(k_1,\tau_{\rm em})
\alpha(k_2,\tau_{\rm em}) \vs
&\times B_{\bar\zeta\bar\zeta\bar\zeta}(\vk_L,\vk_1,\vk_2),
\ea
where 
\be
\alpha(k) = \frac53 \frac{2}{3} \frac{k^2 T(k) g(z)}{\Omega_{m} H_0^2 (1+z)}
\label{eq:alpha}
\ee
is the relation in Fourier space between the density and the curvature
perturbation, $T(k)$ is the matter transfer function normalized to unity as $k\to 0$, and $g(z)$ is the linear growth rate of the gravitational potential normalized to unity during the matter dominated epoch.  

In order to derive the scale-dependent bias, we now assume that the tracer abundance is mostly sensitive to the variance of the density field on a scale $R$, typically chosen to correspond to the Lagrangian scale of the tracer.  See \cite{PBSpaper,schmidt:2013} for a more general and detailed discussion.  The
contribution to the tracer two-point function is then proportional to  $\<\d_F(\v{0}) (\d_{F,R})^2(\vr)\>$, where $\d_{F,R}$ denotes the FNC-frame density field smoothed on scale $R$.  As shown in, e.g.~\cite{long,scoccimarro/etal:2012}, the large-scale scale-dependent bias is then given by
\ba
\D b(k_L) =\:& \frac{\partial\ln \bar n_h}{\partial \s_R^2} \alpha^{-1}(k_L) \D\sigma_R^2(k_L) \vs
\D \sigma_R^2(k_L) =\:& \int \frac{d^3 k_1}{(2\pi)^3} 
\alpha(k_1) \alpha(|\vk+\vk_1|) 
\frac{B_{\bar\zeta\bar\zeta\bar\zeta}(k_L,k_1,|\vk+\vk_1|)}{P_{\bar\zeta}(k_L)} \vs
& \hspace*{1cm}\times \tilde W_R(k_1) \tilde W_R(|\vk+\vk_1|),
\ea
where $\tilde W_R(k)$ is the filter function in Fourier space, and we have 
dropped the $\tau_{\rm em}$ arguments for brevity.  As long as $k_L$ is much
less than the typical values of $k_1$ in the integrand ($k_1 \gg 0.01\iMpch$),
the bispectrum here is evaluated in the squeezed limit \cite{biasbeyond}.  Let us perform
a leading order expansion in this limit,
\be
B_{\bar\zeta\bar\zeta\bar\zeta}(k_L,k_1,|\vk+\vk_1|)
= \left(f_0 + f_2 \frac{k_L^2}{k_S^2} \right) P_{\bar\zeta}(k_L) P_{\bar\zeta}(k_S),
\nonumber
\ee
where $k_S$ is as defined before.  Here, we have assumed a scale-invariant 
bispectrum (otherwise, the coefficients $f_0,\,f_2$ are in general functions 
of $k_S$), and no dependence on the angle between $k_L$ and $k_S$ as is typically
the case (but see \cite{solidinflation} for a counterexample).  The lowest order piece $\propto f_0$ leads to a scale dependence of 
\be
\D b(k_L) \propto f_0\: \alpha^{-1}(k_L) \propto f_0\: k_L^{-2} T^{-1}(k_L),
\ee
which corresponds to the usual scale-dependent bias from local non-Gaussianity.  
The second part leads to a scale-dependence of 
\be
\D b(k_L) \propto f_2\: k_L^2\: \alpha^{-1}(k_L) \propto f_2\: T^{-1}(k_L),
\ee
which is much weaker and only relevant on scales $k_L \gtrsim 0.02\iMpch$ 
where the transfer function departs from unity.  
Since $f_0=0$ in single field inflation [\refeq{3ptschem}], we conclude 
there is \emph{no scale-dependent bias induced on scales} $k_L \lesssim 0.02\iMpch$
in these models.  

The scale-dependent bias discussed here refers to the FNC frame of the tracers; there are 
contributions to the observed scale-dependent bias from projection effects, i.e.
from transforming from the tracer FNC frame to observed positions and redshifts (in our own FNC frame).  
Those have been derived in \cite{yoo/etal:2009,challinor/lewis:2011,bonvin/durrer:2011,BaldaufEtal,gaugePk}, but are generally quite small.  Matching to the
scale-dependent bias from local non-Gaussianity, they correspond to $\fNL < 1$
for a wide range of tracer parameters \cite{gaugePk}.

\section{Conclusions}
\label{sec:concl}

In this work we have presented a simple and complete framework to translate the squeezed limit of any primordial three-point function into observables such as the squeezed limit of the CMB temperature bispectrum and the scale-dependent halo bias. In \reffig{overview} we have spelled out the various steps of the computation and the respective choices of gauge. The different gauges have been chosen in such a way that the whole computation from horizon exit during inflation until observation on earth can be performed using only linear perturbation theory. This required the introduction of a conformal version of the well known Fermi Normal Coordinates (used in \cite{Senatore:2012ya,Senatore:2012wy} for inflationary correlators), in which the spacetime is locally FLRW as opposed to Minkowski. As an example we have applied our formalism to standard single-field slow-roll inflation. We have shown that Maldacena's consistency condition \cite{maldacena:2003} in the squeezed limit of the scalar bispectrum implies that the signal in the CMB bispectrum and halo bias vanishes exactly. Although this result was already known, our approach provides a simple, concise and physically clear derivation.  This calculation can straightforwardly be generalized to higher N-point functions in the squeezed limit.

\chg{In addition, our approach sheds light on the proposal that there are observable correlations between long wavelength tensor modes and short wavelength scalar perturbations \cite{MasuiPen,GiddingsSloth,GiddingsSloth2} (see also \cite{JeongKamionkowski}).  These authors state that the correlation vanishes as long as the long tensor mode is outside of the Hubble radius $k \ll aH$. This can be understood using the results of our section \refsec{fnc}, where we show that a constant and a pure gradient mode of the metric are absorbed by the change of coordinates when going to the Fermi Normal frame. The physical effects of a long mode that a local observer can measure are hence suppressed at least by $(k/aH)^{2}$ and are therefore small for superhorizon perturbations.  Notice in particular that the standard result for the tensor-scalar-scalar bispectrum in single field inflation \cite{maldacena:2003} is just a restatement of this fact but using comoving coordinates. In these coordinates the tensor-scalar-scalar bispectrum takes exactly the right form such that, changing to Fermi Normal coordinates, one finds vanishing correlation up to corrections of order $(k/aH)^{2}$ as we saw in \refsec{comgauge}. After the tensor mode enters the horizon, it starts oscillating and decays. 
During the epoch around horizon crossing, $k\sim aH$, the tensor mode can induce some tidal effects on the short scale scalar power spectrum (see \cite{GWshear} for an evaluation of this effect for the shear).  We did not compute this effect here, and leave this interesting possibility for future work.  
Further, as discussed in \refsec{latetime}, there are projection (photon propagation) effects induced by the long-wavelength tensor mode \cite{KaiserJaffe,DodelsonEtal,GWshear}.  However, these are again suppressed by $(k/H_0)^2$ in the $k\to 0$ limit.  In summary, just as for the scalar bispectrum, a detection of a squeezed-limit tensor-scalar-scalar bispectrum through the measurements described in \cite{MasuiPen,JeongKamionkowski}, at a level larger than expected from tidal and projection effects, would rule out single-field inflation.  }

The framework we have developed in this paper is also useful in other classes of models that have not yet been directly related to observations. One example is resonant non-Gaussianity \cite{Chen:2006xjb,Chen:2008wn,Flauger:2010ja,Leblond:2010yq,Hannestad:2009yx}, which is a generic prediction of many models of axion inflation (see \cite{axionreview} and references therein), in particular inflation from axion monodromy \cite{McAllister:2008hb,Flauger:2009ab,Berg:2009tg}. In these models the inflationary potential has sinusoidal modulations, which lead to oscillations in the time evolution of the background. Since these oscillations average to zero over a period, they can give corrections to the slow-roll parameter that are larger than is usually allowed and nevertheless be perfectly compatible with current power spectrum data (see e.g.~\cite{Flauger:2009ab,Aich:2011qv,Peiris:2013opa}). The resonant bispectrum satisfies Maldacena's consistency condition \cite{Flauger:2010ja,Creminelli:2011rh}, but because of oscillations the amplitude of the primordial bispectrum in the squeezed limit is not suppressed by small slow-roll corrections. In light of this, one might wonder whether these models lead to some detectable signal in the scale dependent halo bias.  \citet{CyrRacine:2011rx} showed that indeed resonant non-Gaussianity produces oscillations in the mass dependence of the non-Gaussian halo bias, which is a very unique signature.  However, the leading contribution in the large-scale limit to the effect derived in that paper is only a coordinate artefact as shown here.  While the halo bias thus has to asymptote to a scale-independent value in the large-scale limit, we expect some interesting effects on intermediate scales.  To compute the actual size of this effect it is important to re-write the primordial resonant bispectrum in terms of FNC. It would be very interesting to perform this analysis using the framework constructed in this paper. There are also other models of the early universe (both inflationary and not) in which the metric perturbations do not freeze outside of the horizon. For example, it has been argued in \cite{Namjoo:2012aa,Chen:2013aj} (see also \cite{Kinney:2005vj}) that this allows one to violate Maldacena's consistency condition. Another interesting model with a peculiar behavior in the squeezed limit is Khronon inflation, studied in \cite{Creminelli:2012xb}. It would be interesting to use our approach to derive the observational prediction of these and similar models.

On the other hand, multifield inflationary models in general feature a non-trivial bispectrum in the squeezed limit.  In this case, curvature perturbations evolve outside the horizon, and the small-scale fluctuations are sensitive to their presence as they essentially evolve in a different FRW background (or, more generally, in a homogeneous anisotropic Universe).  This reasoning can be used to connect the squeezed-limit correlators in multifield inflation to late-time observables in a similar way as outlined here for single-field inflation.


\section*{Acknowledgments}

We acknowledge useful discussions with Liang Dai, Raphael Flauger, Steven Giddings, Donghui Jeong, Marc Kamionkowski, Justin Khoury, Kiyoshi Masui, Ue-Li Pen, and Martin Sloth.  E.~P.~is supported in part by the Department of Energy grant DE-FG02-91ER-40671.  F.~S. is supported by NASA through Einstein Postdoctoral Fellowship grant number PF2-130100 awarded by the Chandra X-ray Center, which is operated by the Smithsonian Astrophysical Observatory for NASA under contract NAS8-03060.  M.~Z.~is supported in part by the National Science Foundation grants PHY-0855425, AST-0907969, PHY-1213563 and by the David \& Lucile Packard foundation.


\appendix


\section{Coordinate transformation of two-point correlations}
\label{app:corrgen}

Consider a scalar field $\rho(x) = \rho(\vx,\tau)$.  Under a general coordinate
transformation $x \to x'(x)$, $\rho$ transforms as
\be
\rho'(x') = \rho(x(x'))\,.
\ee
We define the unequal-time two-point correlation of $\rho$, measured within a patch $p$ through
\ba
& \< \rho(\vx,\tau_1) \rho(\vx+\vr,\tau_2) \>_p  \vs
&\quad = \left\<\, \left[ \rhob(\tau_1) + \d\rho(\vx,\tau_1) \right] \left[ \rhob(\tau_2) + \d\rho(\vx+\vr,\tau_2) \right]\,\right\>_p \vs
& \quad = \rhob(\tau_1) \rhob(\tau_2) + \xi(\vr,\tau_1,\tau_2)\,,
\label{eq:xirho}
\ea
where a subscript $p$ indicates an average over the patch, and we have defined $\rhob(\tau) = \< \rho(\vx,\tau) \>_p$, so that by construction the average over the patch of $\d\rho(\vx,\tau)$ vanishes.  The correlation function introduced above is then defined as
\be
\xi(\vr,\tau_{1},\tau_{2}) \equiv \< \d\rho(\vx,\tau_{1}) \d\rho(\vx+\vr,\tau_{2}) \>_p\,.
\ee
We will consider an alternative definition, the correlation function of $\d\rho/\rho$, below and show that it transforms in the same way.  Further, we will denote the equal time correlation function as $\xi(\vr,\tau)$ in the following for simplicity.  

Now let us consider the equal-time two-point correlation for $\rho'$, measured in the same patch:
\ba
\< \rho'(\vx'_1,\tau') \rho'(\vx'_2,\tau') \>_p =\:& \left\< \rho(\vx_1,\tau_1) \rho(\vx_2,\tau_2) \right \>_p\,,
\ea
where $\vx'_2-\vx'_1 = \vr'$, and we have defined for convenience $\vx_1 \equiv \vx(\vx_1',\tau'),\:\vx_2\equiv \vx(\vx_2', \tau')$, and analogously for $\tau_1,\tau_2$.  By assumption, the coordinate transform is slowly varying over the patch.  We then Taylor expand the coordinate transformations around the center point $\vx'_0 = (\vx'_1+\vx'_2)/2$, 
\ba
x_1^i =\:& x_0^i - \frac12 \frac{\partial x^i(\vx'_0,\tau')}{\partial x'^j} r'^j \,; \quad
x_2^i = x_0^i + \frac12 \frac{\partial x^i(\vx'_0,\tau')}{\partial x'^j} r'^j \vs
\tau_1 =\:& \tau_0 - \frac12 \frac{\partial \tau(\vx'_0,\tau')}{\partial x^j} r'^j \,;\quad
\tau_2 = \tau_0 + \frac12 \frac{\partial \tau(\vx'_0,\tau')}{\partial x^j} r'^j \,,
\label{eq:texp}
\ea
where $\vx_0 = \vx(\vx'_0,\tau')$, and $\tau_0 = \tau(\vx'_0,\tau')$.  We will
discuss below why going to linear order in $r'$ is sufficient.  We then have
\ba
& \< \rho'(\vx'_1,\tau') \rho'(\vx'_2,\tau') \>_p  \vs
&= \rhob(\tau_1)\rhob(\tau_2) + \xi\left(\frac{\partial x^i(\vx_0',\tau')}{\partial x'^j} r'^j, \tau_1, \tau_2
\right) \vs
& = \left[1 + \frac12 \frac{\partial \tau(\vx'_0,\tau')}{\partial x^j} r'^j \frac{\partial}{\partial\tau_1}
- \frac12 \frac{\partial \tau(\vx'_0,\tau')}{\partial x^j} r'^j
\frac{\partial}{\partial\tau_2}
\right] \vs
&\qquad \times\left[\rhob(\tau_1) \rhob(\tau_2) + \xi\left(\frac{\partial x^i(\vx_0',\tau')}{\partial x'^j} r'^j,\tau_1, \tau_2\right)\right]_{\tau_1=\tau_2=\tau_0} \vs
& = \rhob^2(\tau_0) + \xi\left(\frac{\partial x^i(\vx_0',\tau')}{\partial x'^j} r'^j,\tau_0 \right)\,,
\ea
since $\xi(\vr,\tau_1,\tau_2)$ is symmetric in $\tau_1$ and $\tau_2$.  
Thus, at linear order in $r'$, the equal time correlator in primed
coordinates is an equal time correlator in unprimed coordinates.  We thus have
\ba
\< \rho'(\vx',\tau') \rho'(\vx'+\vr',\tau') \>_p =\:& \rhob^2(\tau_0) + \xi(A \vr', \tau_0) \vs
A^i_{\  j} =\:& \frac{\partial x^i(\vx'_0,\tau')}{\partial x'^j}\,.
\label{eq:xirhop1}
\ea
We now define the correlation function of $\rho'$ in the same way as that for $\rho$ [\refeq{xirho}], which yields
\ba
\< \rho'(\vx',\tau') \rho'(\vx'+\vr',\tau') \>_p =\:& \rhob'^2(\tau') + \xi'(\vr',\tau')\,,
\label{eq:xirhop2}
\ea
where 
\be
\rhob'(\tau') = \< \rho'(\vx',\tau') \>_p = \< \rho(x(\vx',\tau')) \>_p = \rhob(\tau_0)\,.
\label{eq:rhobprime}
\ee
The second equality again holds to linear order in $\vx'-\vx'_0$, at which order the average of $\tau$ over the patch at fixed $\tau'$ is just $\tau_0$.  We will discuss this definition of $\rhob'$ in \refapp{bgtransf} below.  
Thus, the first term in each of \refeqs{xirhop1}{xirhop2} agrees, and hence so must the second term:
\be
\xi'(\vr',\tau') = \xi(\vx_2-\vx_1,\tau_0) = \xi\left(A \vr', \tau(\vx_0',\tau') \right)\,.
\label{eq:xitransgen}
\ee
We can similarly derive what happens to $\tilde \xi = \xi / \rhob^2$, which is
the correlation function defined in terms of $\d\rho/\rhob$.  In this case,
we have
\ba
\< \rho(\vx,\tau) \rho(\vx+\vr,\tau) \> =\:& \rhob^2(\tau) \left[ 1 + \tilde\xi(\vr,\tau) \right]\,.
\label{eq:xirhot}
\ea
For $\rho'$, this yields
\ba
\< \rho'(\vx',\tau') \rho'(\vx'+\vr',\tau') \>_p =\:& \rhob^2(\tau_0)\left[1 + \tilde\xi(\vx_2-\vx_1, \tau_0)\right]\,.
\label{eq:xirhotp1}
\ea
Now we define the correlation function of $\rho'$ in the same way as that for $\rho$ [\refeq{xirho}], which yields
\ba
\< \rho'(\vx',\tau') \rho'(\vx'+\vr',\tau') \>_p =\:& \rhob'^2(\tau')\left[1 + \tilde\xi'(\vr',\tau')\right]\,.
\label{eq:xirhotp2}
\ea
Thus, we see that $\tilde\xi$ transforms in the same way as $\xi$, namely
\be
\tilde\xi'(\vr',\tau') = \tilde\xi(\vx_2-\vx_1,\tau_0) = \tilde\xi\left(A \vr', \tau(\vx_0',\tau') \right)\,.
\label{eq:xitransgent}
\ee
There are thus two ways in which the general affine coordinate transform affects the correlation function in primed coordinates: first, there is the spatial transformation of the separation vector, $\vr \to A \vr'$; second, the primed correlation function is evaluated at a different point in space and time (here, we have only made the time shift explicit in the notation, but note that the corrrelation function on the right hand side is to be evaluated within a patch centered on $\vx(\vx'_0, \tau')$).

The linear order expansion in \refeq{texp} neglects higher derivative terms
in the coordinate transformation, i.e.~it is valid in the limit that the
transformation is slowly varying over the patch.  The terms we are neglecting 
in \refeq{texp} correspond to terms of order $k_L^2$ and higher, where $k_L$ 
is the wavenumber of the long-wavelength mode that contributes to the
coordinate transformation.  Since we are neglecting terms of order $k_L^2$
throughout this paper, it is sufficient to expand to linear order.  

A second simplification occurs because we are only considering three-point
correlations in this paper.  In this case, it is sufficient to consider the
linear response of $\xi'(\vr',\tau')$ to the coordinate transformation.  We thus
write
\ba
A^i_{\  j} =\:& \d^i_{\  j} - a^i_{\  j} \vs
\D\tau =\:& \tau(\vx'_0,\tau') - \tau'\,
\ea
and work to linear order in $a^i_{\  j}$ and $\D\tau$.  Further, we can neglect
the effect of the spatial shift in the position at which $\xi$ is evaluated.  This is because the contribution from this shift,
\be
\xi'(\vr',\tau') \supset [\vx(\vx'_0,\tau') - \vx'_0]\cdot \nabla_{\vx'_0} \xi(\vr',\tau')|_{\vx'_0}
\ee
is second order, since both the spatial shift and the location dependence 
of $\xi$ are at least linear order in the long-wavelength perturbations.  \refeq{xitransgen} then simplifies to \refeq{xitranslin}, 
\be
\xi'(\vr';\tau') = \left[1 - a_{ij} r'^i \partial_{r'}^j + \D\tau\,\partial_\tau\right] \xi(\vr';\tau')\,.
\label{eq:xitranslinA}
\ee

 
\subsection{Coordinate transformations in the presence of non-trivial backgrounds}
\label{app:bgtransf}

There is a conceptual point to clarify about the transformation properties of perturbations under changes of coordinates. There are two different but equivalent ways to work with coordinate transformations when studying perturbations on top of some non-trivial background. The first one consists in changing every field in the same way, whether it is going to be treated as a background or not in the rest of the computation. This ``democratic'' approach is conceptually very simple. For example, if a scalar field transforms as $\rho'(x')=\rho(x(x'))$, so does its background $\bar\rho'(x')=\ex{\rho'(x')}=\ex{\rho(x)}=\bar \rho (x)$. Using this point of view, perturbation of the scalar field around that same background must transform in the same way as well. To see this, consider $\d (x)\equiv \rho(x)-\bar \rho(x)$ (notice that $\ex{\d}=0$). Then
\ba
\d'(x')=\rho'(x')-\bar\rho'(t')= \rho(x)-\bar\rho(t)=\d(x)\,,
\ea
In the second approach to the problem one splits the field in perturbation and background in such a way that the latter is invariant under coordinate transformations, i.e.~$\bar\rho'(x)=\bar\rho(x)$ (in the cosmological context, the background quantity is only a function of $\tau$).  This simplification comes at the cost of more complicated transformation laws for the perturbations, which now do \textit{not} transform as a scalar field. Considering the same example as above and a transformation $x'=x+\epsilon$, at linear order in $\epsilon$ one gets
\bea\label{eq:vici}
\d'(x')& =&\rho'(x')-\bar \rho'(x')=\rho'(x')-\bar \rho(x') \\
&=&\rho(x)-\bar \rho(x)+\epsilon^{\mu}\partial_{\mu}\bar\rho(x)= \d(x)+\epsilon^{\mu}\partial_{\mu}\bar\rho(x)\,.\nonumber
\eea
Notice that in both approaches $\rho'(x')=\rho(x(x'))$, while the two different conventions above concern only the transformations of background and perturbations. A similar discussion can be given for tensor fields such as the metric. Again one has two choices: either one defines metric perturbations that transform like the component of a tensor, in which case the metric background changes when one changes coordinates or the background is assumed fixed in any coordinates and metric perturbations have additional terms in their transformations. 

We warn the reader that in the derivation of \refapp{corrgen} we have used the ``democratic convention'' as opposed to the fixed background convention which is standard in the cosmological literature (e.g.~\cite{Weinberg:2008zzc}).  However, it should be kept in mind that around a homogeneous background with a long wavelength perturbation, the additional contribution in the last term of \refeq{vici} is just a constant or a pure gradient at the order we are working in, and hence does not contribute to the correlation function evaluated on much smaller scales $r$.  This means that our final result \refeq{xitranslinA} is valid independently of the convention used for defining the perturbations.

\begin{widetext}
\section{Squeezed-limit three-point function from transformed two-point correlation}
\label{app:sqlimit}

This section derives the squeezed-limit three-point function and 
bispectrum from a coordinate transformation of the two-point function
given by \refeq{xitranslinA},
\be\label{eq:xitranslin2}
\xi'(\vr;\tau) = \left[1 - a_{ij} r^i \partial_{r}^j + \D\tau\,\partial_\tau\right] \xi(\vr;\tau)\,,
\ee
where we have dropped the prime on coordinates since we will only deal
with primed coordinates in this section.  Specifically, we want to
derive the three point function $\<X'(\vx_3) \d'(\vx_1) \d'(\vx_2) \>$ in the
limit where $|\vx_3 - \vx_1| \gg |\vx_1-\vx_2|$ (squeezed limit).  Here, $X$ and $\d$ stand for
any perturbation variables with mean zero (of course, the auto-three-point function $X=\d$ is a special case).  In this limit, the three-point function 
quantifies the modulation of the local two-point function $\xi'(|\vx_1-\vx_2|)$
by a long-wavelength perturbation $X'$ (modes with wavelength much less than
$|\vx_3-\vx_1|$ will not contribute to this correlation).  As discussed in \refsec{sqlimit}, the coordinate transformation only acts on the small-scale fluctuations, so that we set $X'(\vx) = X(\vx)$.  Thus,
\ba
\<X'(\vx_3) \d'(\vx_1) \d'(\vx_2) \> \stackrel{\rm squeezed}{=}\:&
\< X'(\vx_3) \xi'(\vx_1-\vx_2;\tau)|_{X'(\vx_c)}\>,
\label{eq:XddA}
\ea
where we have defined the point of evaluation of the long-wavelength $X$ perturbation as
\be
\vx_c = c \vx_1 + (1-c) \vx_2\,.
\ee
Choosing $\vx_c$ along the axis connecting $\vx_1$ and $\vx_2$ is necessary since \refeq{XddA} describes a homogeneous and isotropic three-point function.  The midpoint $\vx_0 = (\vx_1+\vx_2)/2$ corresponds to $c = 1/2$.  We
will see that the choice of $c$ does not influence the final result below.  
Using \refeq{xitranslin2}, this yields
\ba
\<X'(\vx_3) \d'(\vx_1) \d'(\vx_2) \> \stackrel{\rm squeezed}{=}\:&
\left[ - \xi_{X a_{ij}}(|\vx_3-\vx_c|) r^i \partial_{r}^j + \xi_{X \D\tau}(|\vx_3-\vx_c|)\,\partial_\tau\right] \xi(\vx_1-\vx_2;\tau)
+ \< X(\vx_3) \xi(\vx_1-\vx_2; \tau)|_{X(\vx_c)}\> \vs
=\:& \left[ - \xi_{X a_{ij}}(|\vx_3-\vx_c|) r^i \partial_{r}^j + \xi_{X \D\tau}(|\vx_3-\vx_c|)\,\partial_\tau\right] \xi(\vx_1-\vx_2;\tau)
+ \< X(\vx_3) \d(\vx_1) \d(\vx_2) \>\,.
\label{eq:xi3sq1}
\ea
This expression gives the squeezed-limit three-point function of $X'$ and $\d'$ (in the \emph{primed} coordinates) in terms of derivatives of $\xi(\vr,\tau)$ and the three-point function of $X$ and $\d$ in the \emph{unprimed} coordinate frame.

The left-hand side of \refeq{xi3sq1} is clearly symmetric under $\vx_1\leftrightarrow \vx_2$.  For a general choice of $\vx_c$, this does not hold for the right-hand side, so we should symmetrize:
\ba
\<X'(\vx_3) \d'(\vx_1) \d'(\vx_2) \> \stackrel{\rm squeezed}{=}\:&
\frac12\bigg[ - \xi_{X a_{ij}}(|\vx_3-\vx_c|) r^i \partial_{r}^j
- \xi_{X a_{ij}}(|\vx_3-\vx'_c|) r^i \partial_{r}^j \vs
& \quad + \xi_{X \D\tau}(|\vx_3-\vx_c|)\,\partial_\tau
+ \xi_{X \D\tau}(|\vx_3-\vx'_c|)\,\partial_\tau\bigg] \xi(\vx_1-\vx_2;\tau) 
+ \< X(\vx_3) \d(\vx_1) \d(\vx_2) \>\,,
\label{eq:xi3sq2}
\ea
where $\vx'_c = (1-c) \vx_1 + c \vx_2$.  Of course, if we choose $\vx_c$
to be the midpoint of $\vx_1$ and $\vx_2$, the two permutations are identical.  

We now derive the Fourier-space analog of \refeq{xi3sq2}.  In terms of the
cross-power spectra $P_{X a_{ij}}(k),\:P_{X \D\tau}(k)$ and the auto power
spectrum $P_\d(k)$, we have
\ba
\< X(\vx_3) a_{ij}(\vx_c)\> =\:& \int \frac{d^3\vk_L}{(2\pi)^3}
P_{X a_{ij}}(k_L) e^{i \vk_L [\vx_{3} - \vx_c]} 
\vs
\< X(\vx_3) \D\tau(\vx_c)\> =\:& \int \frac{d^3\vk_L}{(2\pi)^3}
P_{X \D\tau}(k_L) e^{i \vk_L [\vx_{3} - \vx_c]} 
\vs
\bar r^j \partial_i \xi(\vr,\tau) =\:& \int \frac{d^3 \vk_S}{(2\pi)^3}
P_\d(k_S,\tau) \bar r^j \partial_i e^{i\vk_S \vr} \vs
=\:& -\int \frac{d^3 \vk_S}{(2\pi)^3}
\frac{\partial}{\partial k_S^j} \left[ k_S^i P_\d(k_S)\right] e^{i\vk_S \vr} \vs
\partial_{\bar\tau} \xi(\bar\vr, \tau) =\:& \int \frac{d^3 \vk_S}{(2\pi)^3}
\partial_\tau P_\d(k_S, \tau) e^{i\vk_S \vr} \,.
\ea
The bispectrum in the squeezed limit $k_L \ll k_1,\,k_2$ becomes
\ba
\< X'(\vk_L) \d'(\vk_1) \d'(\vk_2)\> =\:& \int d^3 \vx_{3}
\int d^3 \vx_{1} \int d^3 \vx_{2} e^{-i(\vk_L \vx_{3}
+ \vk_1 \vx_{1} + \vk_2 \vx_{2})} 
\< X(\vx_{3}) \d'(\vx_{1}) \d'(\vx_{2}) \> \vs
\stackrel{\rm squeezed}{=}\:&
\frac12 \Bigg\{ 
\left[P_{X a_{ij}}(k_L) \frac{\partial}{\partial k_S^j}
\left[ k_S^i P_\d(k_S) \right] + P_{X\D\tau}(k_L) \frac{\partial}{\partial\tau} P_\d(k_S) \right]_{\vk_S = \vk_1+c\vk_L}  + (c \to 1-c) \Bigg\} \vs
& \quad\times (2\pi)^3 \d_D\left(\vk_1 + \vk_2 + \vk_L\right) + \< X(\vk_L) \d(\vk_1) \d(\vk_2)\>\,.
\ea
Thus, the transformed bispectrum has the proper delta function ensuring the triangle condition,
and we can identify the $X' \d' \d'-$bispectrum in the squeezed limit as
\ba
B_{X'\d'\d'}(\vk_L,\vk_1,\vk_2) \stackrel{\rm squeezed}{=}\:&
\frac12 \Bigg\{ 
\left[ P_{X a_{ij}}(k_L) \frac{\partial}{\partial k_S^j}
\left[ k_S^i P_\d(k_S) \right] + P_{X\D\tau}(k_L) \frac{\partial}{\partial\tau} P_\d(k_S) \right]_{\vk_S = \vk_1+c\vk_L} \hspace*{-1cm} + (c \to 1-c) \Bigg\} \vs
& + B_{X\d\d}(\vk_L,\vk_1,\vk_2)\vs
=\:& \frac12 \Bigg\{ 
\left[ P_{X a_{ij}}(k_L) \left(\d_{ij} + \frac{k_S^i k_S^j}{k_S^2} \frac{d}{d\ln k_S} \right) P_\d(k_S) + P_{X\D\tau}(k_L) \frac{\partial}{\partial\tau} P_\d(k_S) \right]_{\vk_S = \vk_1+c\vk_L} \hspace*{-1cm} + (c \to 1-c) \Bigg\} \vs
& + B_{X\d\d}(\vk_L,\vk_1,\vk_2)\,.\label{eq:Bsq1}
\ea
Since we are working in the squeezed limit, we can expand in $k_1/k_L$.  
We have
\ba
F(|\vk_1+ c \vk_L|) =\:& F(k_1) + c \hat\vk_1\cdot\hat\vk_L \frac{k_L}{k_1} \frac{d}{d\ln k_1} F(k_1) + \O(k_L^2/k_1^2) \vs
\frac{k_S^i k_S^j}{k_S^2} =\:& \frac{k_1^i k_1^j}{k_1^2} \left(1 - 2c \frac{\vk_1\cdot\vk_L}{k_1^2}\right) + 2 c \frac{k_1^{(i} k_L^{j)}}{k_1^2} + \O(k_L^2/k_1^2) \vs
\frac{d P_\d(k_s)}{d\ln k_S} =\:& \frac{d P_\d(k_1)}{d\ln k_1} + c \hat\vk_1\cdot\hat\vk_L \frac{k_L}{k_1} \frac{d^2}{d(\ln k_1)^2} P_\d(k_1) + \O(k_L^2/k_1^2) \,.
\ea
We already see that all corrections are linear in $c$, so that the two
permutations $c,\:1-c$ add up to 1.  Thus, the result becomes independent of the choice of $c$, and the squeezed limit bispectrum in terms is, to order $(k_L^2/k_1^2)$,
\ba
B_{X'\d'\d'}(\vk_L,\vk_1,\vk_2) \stackrel{\rm squeezed}{=}\:&
P_{X a_{ij}}(k_L) \left(\d_{ij} + \frac{k_1^i k_1^j}{k_1^2} \frac{d}{d\ln k_1} \right) P_\d(k_1) + P_{X\D\tau}(k_L) \frac{\partial}{\partial\tau} P_\d(k_1) \vs
& + \hat\vk_1\cdot\hat\vk_L \frac{k_L}{k_1} \frac12 \Bigg\{ 
 P_{X a_{ij}}(k_L) \left(\d_{ij} +  \frac{k_1^i k_1^j}{k_1^2} \left[\frac{d}{d\ln k_1} -2 \right]\right) \frac{d P_\d(k_1)}{d\ln k_1} + P_{X\D\tau}(k_L) \frac{\partial}{\partial\tau} \frac{dP_\d(k_1)}{d\ln k_1} \Bigg\} \vs
& + P_{X a_{ij}}(k_L) \frac{k_1^{(i} k_L^{j)}}{k_1^2} \frac{dP_\d(k_1)}{d\ln k_1} 
+ B_{X\d\d}(\vk_L,\vk_1,\vk_2)\,.\label{eq:Bsq2}
\ea
Since $\vk_2 = -\vk_1 + \O(k_L/k_1)$, this expression can be equivalently written in terms of $\vk_2$ instead of $\vk_1$.  Since the choice of $c$ is arbitrary, we will use the most natural choice, $c=1/2$, in which case \refeq{Bsq1} becomes simply
\ba
B_{X'\d'\d'}(\vk_L,\vk_1,\vk_2) \stackrel{\rm squeezed}{=}\:&
\left[P_{X a_{ij}}(k_L) \left(\d_{ij} + \frac{k_S^i k_S^j}{k_S^2} \frac{d}{d\ln k_S} \right) P_\d(k_S) + P_{X\D\tau}(k_L) \frac{\partial}{\partial\tau} P_\d(k_S) \right]_{\vk_S = \vk_1+\vk_L/2} \vs
& + B_{X\d\d}(\vk_L,\vk_1,\vk_2)\,.\label{eq:Bsq3}
\ea
While this expression is equivalent to \refeq{Bsq2} up to order $k_L^2$, we will work with this result since it is more compact and convenient.

Including the effect of a mean density modulated by $X$ following \refeq{ctransf} in \refsec{latetime} is now an obvious generalization.  We obtain
\ba
B_{X'\d'\d'}(\vk_L,\vk_1,\vk_2) \stackrel{\rm squeezed}{=}\:&
\left[P_{X a_{ij}}(k_L) \left(\d_{ij} + \frac{k_S^i k_S^j}{k_S^2} \frac{d}{d\ln k_S} \right) P_\d(k_S) + P_{X\D\tau}(k_L) \frac{\partial}{\partial\tau} P_\d(k_S) + 2 P_{X c}(k_L) P_\d(k_S)\right]_{\vk_S = \vk_1+\vk_L/2} \vs
& + B_{X\d\d}(\vk_L,\vk_1,\vk_2)\,.\label{eq:Bsqfinal}
\ea
We can now further decompose $a_{ij}$ as
\be
a_{ij} = \frac1{n_D} a \d_{ij} + a^T_{ij},
\ee
where $n_D$ is the dimensionality of the space in which we define the
correlations, and $a_{ij}^T$ is traceless.  In particular $n_D=3$ for 
three-dimensional observables
such as galaxy densities or 21cm flux, and $n_D=2$ for projected quantities
on the sky such as the CMB.  We then obtain
\ba
B_{X'\d'\d'}(\vk_L,\vk_1,\vk_2) \stackrel{\rm squeezed}{=}\:&
\bigg[P_{X a}(k_L) \left(1 + \frac1{n_D} \frac{d}{d\ln k_S} \right) P_\d(k_S) 
+ P_{X a^T_{ij}}(k_L) \frac{k_S^i k_S^j}{k_S^2} \frac{dP_\d(k_S)}{d\ln k_S} \vs
& \quad + P_{X\D\tau}(k_L) \frac{\partial}{\partial\tau} P_\d(k_S) + 2 P_{X c}(k_L) P_\d(k_S)\bigg]_{\vk_S = \vk_1+\vk_L/2} + B_{X\d\d}(\vk_L,\vk_1,\vk_2)\vs
=\:& \bigg[P_{X a}(k_L) P_\d(k_S) \frac{d\ln (k_S^{n_D} P_\d(k_S))}{d\ln k_S}
+ P_{X a^T_{ij}}(k_L) \frac{k_S^i k_S^j}{k_S^2} \frac{dP_\d(k_S)}{d\ln k_S} \vs
& \quad + P_{X\D\tau}(k_L) \frac{\partial}{\partial\tau} P_\d(k_S) + 2 P_{X c}(k_L) P_\d(k_S)\bigg]_{\vk_S = \vk_1+\vk_L/2} + B_{X\d\d}(\vk_L,\vk_1,\vk_2)
\,.\label{eq:Bsqfinald}
\ea
Note that the contribution of the trace of $a_{ij}$ scales as the logarithmic deerivative of $k^{n_D} P_\d(k)$, while the trace-free component $a^T_{ij}$ couples to the logarithmic derivative of $P_\d(k)$ itself.

\end{widetext}


\section{Fermi Normal Coordinates}
\label{app:FNC}

\subsection{FNC construction and metric}

In this section, we show how for a general metric, the coordinates
$x_F^\mu$ defined through \refeq{FNCcoordTF} below in fact do lead
to a metric of the form
\be
g_{\mu\nu}^F = \eta_{\mu\nu} + \O(x_F^2) \, .
\label{eq:gmunuF}
\ee
To quadratic order, the Fermi coordinate transformation is given by
\ba
x^\mu(x_F^i)\Big|_{t_P}
= x^\mu(P) +
\left(e_i\right)^\mu_P x_F^i
-
\frac{1}{2}
\Gamma^\mu_{\alpha\beta}\Big|_P
(e_i)^\alpha_P (e_j)^\beta_P
x_F^ix_F^j
.\label{eq:FNCcoordTF}
\ea
Straightforward algebra yields the transformation matrix as
\ba
\frac{\partial x^\alpha}{\partial x_F^\mu} = (e_\mu)_P^\alpha + A^\alpha_{\mu j} x_F^j + \O(x_F^2),
\label{eq:dxdxF}
\ea
where a subscript $P$ denotes the evaluation at the point on the central
geodesic specified by $x_F^0$.  Here,
\ba
A^\alpha_{0j} =\:& (e_0)^\beta_P \left[\partial_\beta (e_j)^\alpha\right]_P
\vs
A^\alpha_{ij} =\:& - \Gamma^\alpha_{\beta\gamma}\Big|_P (e_i)^\beta_P (e_j)^\gamma_P
\ea
Using the fact that the unit vectors $(e_j)^\alpha$ are assumed to be
parallel-transported along the central geodesic, we have
\be
0 = (e_0)^\beta_P \left[\nabla_\beta (e_j)^\alpha\right]_P
= (e_0)^\beta_P \left\{\partial_\beta (e_j)^\alpha
+ \Gamma_{\beta \lambda}^\alpha (e_j)^\lambda \right\}_P,
\nonumber
\ee
which can be used to bring $A^\alpha_{\mu j}$ into a uniform expression:
\be
A^\alpha_{\mu j} =\:& - \Gamma^\alpha_{\beta\gamma}\Big|_P (e_\mu)^\beta_P (e_j)^\gamma_P\,.
\label{eq:Aamj}
\ee
Next, we expand
\be
g_{\alpha\beta}(x(x_F)) = g_{\alpha\beta}\Big|_P + (e_j)_P^\gamma \left[\partial_\gamma g_{\alpha\beta}\right]_P x_F^j.
\ee
Finally, using that 
\be
g_{\alpha\beta}\Big|_P (e_\mu)^\alpha_P (e_\nu)^\beta_P = \eta_{\mu\nu}
\ee
by construction of the orthonormal tetrad (note that this holds for any 
velocity $v_i$ at order $v$), we obtain
\ba
& g_{\mu\nu}^F(x_F) = g_{\alpha\beta}(x(x_F) )\frac{\partial x^\alpha}{\partial x_F^\mu} \frac{\partial x^\beta}{\partial x_F^\nu} \vs
& = \eta_{\mu\nu} + \left[ \partial_\gamma g_{\alpha\beta}
- 2 g_{\lambda\beta} \Gamma^\lambda_{\alpha\gamma} \right]
(e_{(\mu})^\alpha_P (e_{\nu)})_P^\beta (e_j)^\gamma_P x_F^j \vs
& \quad + \O(x_F^2).
\ea
Assuming that we have a metric connection so that
\be
\nabla_\gamma g_{\alpha\beta} = \partial_\gamma g_{\alpha\beta}
- 2\Gamma^\lambda_{\gamma(\alpha} g_{\lambda \beta)} = 0,
\ee
we see that the order $x_F^j$ correction to the metric vanishes. 

So far, we have not used the condition that the central curve whose
tangent vector is $(e_0)^\mu$ is a geodesic.  However, if we want
the metric in FNC to be of the form \refeq{gmunuF} at more than just one
point along the central curve, we clearly need at lowest order
\be
(e_0)^\gamma \nabla_\gamma \left[g_{\alpha\beta} (e_\mu)^\alpha (e_\nu)^\beta
\right]_P = 0.
\ee
This in particular implies
\be
(e_0)^\gamma \nabla_\gamma (e_0)^\mu = 0
\ee
along the curve,
which is precisely the condition for the central curve to be a geodesic.  

\subsection{Transformation from general coordinates to \fncb}
\label{app:FNCgen}

We now explicitly derive the transformation \refeqs{FNCbtrans}{FNCbtransI} into the \fncb~frame.  We write the conformal metric as
\be
\bar g_{\mu\nu} = \eta_{\mu\nu} + h_{\mu\nu}\,,
\ee
and work to linear order in $h_{\mu\nu}$.  The tetrad around $P$ is 
given by
\ba
(e_0)^\mu =\:& \left(1 + \frac12 h_{00}, \: v^i \right) \vs
(e_j)^\mu =\:& \left(v_j + h_{0j},\: \d_j^{\  i} - \frac12 h_j^{\  i}
\right)\,.
\ea
Note that we treat $v$ as $\O(h)$ here, as we are assuming comoving observers. 
Since $(e_\alpha)^\mu = \d_\alpha^{\  \mu} + \O(h)$ and $\Gamma^{\mu}_{\alpha\beta} = \O(h)$, we have for the
transformation into \fncb~at lowest order in $h$:
\ba
x^\mu(\bar x^\alpha_F) =\:& P^\mu + (e_i)^\mu_P \bar x_F^i
- \frac12 \Gamma^\mu_{ij}\Big|_P x_F^i x_F^j\,,
\ea
where $P^{\mu}\equiv x^{\mu}(P)$ are the global coordinates of the central geodesic, so that $v^{i}=\partial P^{i}/\partial x^{0}$ and we define analogously $\Pb^{\mu}\equiv\bar x_{F}^{\mu}(P)$ for \fncb.  The conformal proper time $\Pb^{0}$ defines the time coordinate of \fncb, namely $\Pb^{0}=\bar x^{0}_{F}$. Without loss of generality, we can choose the spatial origin so that $\Pb^{i}(\tau_F) = 0$ at some fixed proper time $\tau_F$. To lowest order in $h$ and $v$, the conformal proper time  is related to the global time by 
\be
x^0(P) = \bar x_F^0 + \frac12 \int_0^{\bar x^0_F} h_{00}(\v{0},\tau) d\tau\,.
\ee
The Christoffel symbols are given by
\be
\Gamma^\mu_{ij} = \frac12\left[h^\mu_{\  i,j} + h^\mu_{\  j,i} - h_{ij}^{\  \  ,\mu}
\right]\,,
\ee
where indices are raised and lowered with $\eta_{\mu\nu}$.  Thus,
\ba
x^0(\bar x^\alpha_F) =\:& \bar x^0_F + \frac12 \int_0^{\bar x^0_F} h_{00}(\tau) d\tau 
+ (v_i + h_{0i}) \bar x_F^i \vs
& - \frac14 \left[h^0_{\  i,j} + h^0_{\  j,i} - h'_{ij} \right] \bar x_F^i \bar x_F^j \vs
=\:& \bar x^0_F + \frac12 \int_0^{\bar x^0_F} h_{00}(\tau) d\tau 
+ (v_i + h_{0i}) \bar x_F^i \vs
& + \frac14 \left[h_{0i,j} + h_{0j,i} + h'_{ij} \right] \bar x_F^i \bar x_F^j \\
x^k(\bar x^\alpha_F) =\:& P^k - \frac12 h^k_{\  i} \bar x_F^i
- \frac14 \left[h^k_{\  i,j} + h^k_{\  j,i} - h_{ij}^{\  \  ,k}
\right] \bar x_F^i \bar x_F^j\,.
\ea
Here all terms linear in perturbations, namely $h_{\mu\nu}$ and $v^{i}$, should be evaluated along the central geodesic $P^{\mu}$.  For $\bar x^0_F \neq \tau_F$, the spatial position of the geodesic differs from the origin in the $x^\mu$ coordinate system by an amount $\Delta x^{0}_{F} v^{i}$, which is linear in perturbations (we will shortly find $v^{i}$ from the geodesic equation). Therefore, up to terms quadratic in metric perturbations, we can evaluate $h_{\mu\nu}$ and $v^{i}$ at the origin of the FNC $\bar x^{i}_{F}=\Pb^{i}(x_{F\ast}^{0})\equiv 0$.

At linear order in $h$, the inverse transformation is very easily derived:
\ba
\bar x_F^0(x^\alpha) =\:& x^0 - \frac12 \int_0^{x^0} h_{00}(\tau) d\tau 
- (v_i + h_{0i}) x^i \vs
& - \frac14 \left[h_{0i,j} + h_{0j,i} + h'_{ij} \right] x^i x^j \\
\bar x_F^k(x^\alpha) =\:& x^k + \frac12 h^k_{\  i} x^i
+ \frac14 \left[h^k_{\  i,j} + h^k_{\  j,i} - h_{ij}^{\  \  ,k}
\right] x^i  x^j\,.
\ea
Note that when $h_{0i}\neq 0$ there is no solution in which the observer is at rest. We now focus on this case now, where we can set $h_{ij}=h_{00}=0$, since their effect is simply additive and was considered above. For $h_{i0}\neq 0$, one needs to know the relation between $v^{i}$ and $h_{0i}$ that is enforced by the geodesic equation for P. Assuming as we did previously that the velocity $v^{i}=\partial P^{i}/\partial x^{0}$ is of the same order as $h_{0i}$, the geodesic equations at linear order in perturbations in global coordinates are given by
\be
\left(\frac{\partial }{\partial x^{0}}\right)^{2}P^{0}=0\,,\quad \left(\frac{\partial }{\partial x^{0}}\right)^{2} P^{i}+\Gamma^{i}_{00}  \left( \frac{\partial P^{0}}{\partial x^{0}} \right)^{2}=0\,,
\nonumber
\ee
where we used the fact that $\Gamma^{\mu}_{\nu\sigma}$ is at least linear in metric perturbations. The first equation tells us that proper time coincides with the global time coordinate at this order, up to two arbitrary integration constants, which we fix by choosing $P^{0}=x^{0}$. This is no surprise since we are assuming $h_{00}=0$. Then, from the second equation, we find $v^{i}=\partial P^{i}/\partial x^{0}=- h_{0i}$, where we have set another integration constant to zero following the assumption that $\partial_{0} P^{i}$ is linear in $h_{0i}$.   
Specifically, for $\bar x_F^0$ close to $\tau_F$, we have
\be
x^i(P) = v^i (\bar x_F^0 - \tau_F), 
\ee
so that to leading order
\ba
x^k(\bar x^\alpha_F) =\:& v^k (\bar x_F^0 - \tau_F) + \bar x_F^k - \frac12 h^k_{\  i} \bar x_F^i \vs
& - \frac14 \left[h^k_{\  i,j} + h^k_{\  j,i} - h_{ij}^{\  \  ,k}
\right] \bar x_F^i \bar x_F^j
\\
\bar x_F^k(x^\alpha) =\:& x^k - v^k (x^0 - \tau_F)
+ \frac12 h^k_{\  i} x^i \vs
& + \frac14 \left[h^k_{\  i,j} + h^k_{\  j,i} - h_{ij}^{\  \  ,k}
\right] x^i  x^j\,.
\ea

\section{Uniqueness of Fermi Normal Coordinates}
\label{app:FNCU}

In this section we investigate what residual coordinate freedom remains
when requiring that the metric be of ``FNC form'':
\be
g_{\mu\nu}^F = \eta_{\mu\nu} + S_{\mu\nu ij} x_F^i x_F^j.
\label{eq:FNCcond}
\ee
Here, $S_{\mu\nu ij}$ in general depends on the affine parameter along
the central geodesic, i.e.~the Fermi-frame time coordinate $t_F$.  
Let us consider a general inertial frame $\{ x'^\mu \}$ constructed around
point $P$.  Without loss of generality, we let $P$ be at the origin of
both the $\{ x'^\mu \}$ and $\{ x_F^\mu \}$ coordinate systems.  The
requirement that $\{ x'^\mu \}$ be inertial, i.e.~that $g_{\mu\nu}' = \eta_{\mu\nu}$ 
at $P$ and that $\partial'_\alpha g_{\mu\nu}' = 0$ at $P$, restricts the
relation between the coordinates to be at least of cubic order:
\ba
x'^\alpha =\:& x_F^\alpha + \frac13 C^{\alpha}_{\  \beta\gamma\delta} x_F^\beta
x_F^\gamma x_F^\delta + \O(x_F^4) \vs
x_F^\alpha =\:& x'^\alpha - \frac13 C^{\alpha}_{\  \beta\gamma\delta} x'^\beta
x'^\gamma x'^\delta + \O(x'^4).
\ea
Here,
\be
C^{\alpha}_{\  \beta\gamma\delta} = \frac{\partial^3 x'^\alpha}{\partial x_F^\beta x_F^\gamma x_F^\delta}\Big|_P,
\ee
which implies
\be
C^{\alpha}_{\  \beta\gamma\delta} = C^{\alpha}_{\  (\beta\gamma)\delta} 
= C^{\alpha}_{\  \beta(\gamma\delta)} = C^{\alpha}_{\  (\beta\gamma\delta)},
\label{eq:Csymm}
\ee
i.e.~the last equality states that $C^\alpha_{\  \beta\gamma\delta} = C^\alpha_{\  \delta\gamma\beta}$.  
The metric in the primed frame then becomes to order $x'^2$:
\be
g'_{\mu\nu} = \eta_{\mu\nu} + \left[S_{\mu\nu\gamma\delta} - 2 C_{(\mu\nu)\gamma\delta} \right] x'^\gamma x'^\delta.
\ee
By assumption, $S_{\mu\nu\gamma\delta}$ obeys the ``FNC condition''
\be
S_{\mu\nu 0\delta} = 0 = S_{\mu\nu\gamma 0}.
\ee
Imposing the same restriction on $g'_{\mu\nu}$ leads to the condition
\be
C_{(\mu\nu)0 \delta} = 0 = C_{(\mu\nu)\gamma 0}.
\ee
Using the symmetry properties \refeq{Csymm}, we obtain
\be
C_{(\mu 0)\gamma \delta} = 0.
\ee
This says that there is \emph{no freedom} in the order $\O(x_F^2)$ corrections
to the Fermi-frame metric components $g^F_{00}$ and $g^F_{0i}$ for coordinates 
satisfying the condition \refeq{FNCcond} (but see below).  However, there 
is some freedom in choosing the spatial part $g^F_{ij}$, since a non-zero
$C_{ijkl}$ is allowed by \refeq{FNCcond}.  

Finally, there is an additional freedom in the choice of coordinates which
does not affect the metric at order $x_F^2$.  Namely, we can choose
$C_{[0\mu]ij} \neq 0$.  Specifically,
\be
C_{0kij} = - C_{k0ij} = 3\alpha_k A_{ij},
\ee
where we have introduced a 3-vector $\alpha_k$ and a symmetric 3-tensor
$A_{ij}$.  By construction, $C_{(0\mu)\gamma\delta} = 0$.  This corresponds
to a coordinate transform of (note that $C^{0}_{\  kij} = -\alpha_k A_{ij}$)
\ba
x'^0 = x_F^0 - \alpha_k x_F^k \: A_{ij} x_F^i x_F^j \vs
x'^i = x_F^i - \alpha^i x_F^0\: A_{lj} x_F^l x_F^j.
\ea
In other words, this coordinate transform is a spatially location-dependent
Lorentz boost with a velocity given by
\be
v^k(x_F^l) = \alpha^k\: A_{ij} x_F^i x_F^j,
\ee
which vanishes quadratically on the central geodesic.  This type of
coordinate transform leaves the Fermi frame metric entirely invariant at 
order $x_F^2$.  

\bibliography{PBSng}


\end{document}